\documentclass[twocolumn]{autart}    

\usepackage{dsfont, subfigure}
\usepackage{amssymb,mathtools}
\usepackage{array, multirow}
\usepackage{tabularx, graphics, graphicx}
\usepackage{algpseudocode}
\usepackage{booktabs} 
\usepackage{epstopdf}
\usepackage{paralist}
\usepackage[colorlinks=true,linkcolor=black, citecolor=cyan, urlcolor=blue]{hyperref}
\usepackage{marginnote}
\usepackage{color}
\usepackage{apacite} 
\usepackage{algorithm} 
\usepackage{marginnote}

\newtheorem{theorem}{Theorem}
\makeatletter
\renewcommand*{\@opargbegintheorem}[3]{\trivlist
	\item[\hskip \labelsep{\bfseries #1\ #2}] \textbf{(#3)}\ \itshape}
\makeatother
\newtheorem{lemma}{Lemma}
\newtheorem{corollary}{Corollary}
\newtheorem{remark}{Remark}

\newtheorem{definition}{Definition}
\newtheorem{assumption}{Assumption}
\newtheorem{problem}{Problem}

\begin{document}

\begin{frontmatter}

\title{Robust distributed model predictive control of linear systems: analysis and synthesis}


\author{Ye Wang}\ead{ye.wang1@unimelb.edu.au},   
\author{Chris Manzie}\ead{manziec@unimelb.edu.au}               

\address{Department of Electrical and Electronic Engineering, The University of Melbourne, VIC 3010, Australia}  


\begin{abstract}                          
	To provide robustness of distributed model predictive control (DMPC), this work proposes a robust DMPC formulation for discrete-time linear systems subject to unknown-but-bounded disturbances. Taking advantage of the structure of certain classes of distributed systems seen in applications with interagent coupling like vehicle platooning, a novel robust DMPC is formulated. The proposed approach is characterised by separable terminal costs and locally robust terminal sets, with the latter sets adaptively estimated in the online optimisation problem. A constraint tightening approach based on a set-membership approach is used to guarantee constraint satisfaction for coupled subsystems in the presence of disturbances. Under this formulation, the closed-loop system is shown to be recursively feasible and input-to-state stable. To aid in the deployment of the proposed robust DMPC, a possible synthesis method and design conditions for practical implementation are presented. Finally, simulation results with a mass-spring-damper system are provided to demonstrate the proposed robust DMPC.
\end{abstract}

\end{frontmatter}

\section{Introduction}\label{section:introduction}

Model predictive control (MPC) is a powerful methodology~\cite{Maciejowski2002,Rawlings2019} that has been widely considered and used in a variety of industrial applications, such as chemical processes~\cite{Ellis2017}, water networks~\cite{Wang2017} as well as building energy managements~\cite{Oldewurtel2012}. As the size of the system increases, challenges in maintaining a centralised control strategy may arise due to issues including computational complexity and the mixed spatial and/or temporal scales necessitating larger communication requirements, all of which can impact on the ability to meet real-time requirements. 

With the development of communication and distributed optimisation techniques, distributed MPC (DMPC) has been an active research field during the past two decades. DMPC can be also regarded as an alternative way to overcomes issues for centralised control strategy. Recent works on DMPC can be found in the literature \shortcite{Conte2016,Darivianakis2020,maestre2014distributed,Giselsson2014}. Among these DMPC methods, the corresponding DMPC optimisation problem is built in a separable structure so that it can be decomposed into sum of local optimisation problems. Therefore, distributed optimisation techniques, such as dual decomposition~\cite{Farokhi2014}, alternating direction method of multipliers (ADMM) and improved/accelerated  ADMM~\shortcite{Ghadimi2015,Mota2013,Teixeira2016}, and game theoretic approaches~\cite{Julian2019}, can be implemented. Since local optimisation problems are solved in parallel, the computation time is significantly reduced, which is helpful for practical implementation.


For systems subject to disturbances, robustness is necessarily considered in an MPC setup \shortcite{Mayne2005,Rawlings2019}. Tube-based approach as one of popular robust MPC has been widely studied, see e.g. \shortciteA{Alvarado2010}, \shortciteA{Limon2008}, \shortciteA{Broomhead2015}, \shortciteA{Pereira2017}. The idea of tube-based MPC is to optimise the nominal system model along the prediction horizon subject to tightened constraints by means of an effective robust constraint tightening approach. In principle, robust positively invariant (RPI) sets~\shortcite{Stoican2011,Wang2019} are used for constraint satisfaction in order to guarantee recursive feasibility and stability of the closed-loop systems.

In this work we consider  systems subject to unknown but bounded disturbances. Previous research on robust DMPC has considered modifying the controller to guarantee recursive feasibility and closed-loop stability in the presence of disturbances. In \citeA{Richards2007} and \citeA{Trodden2006}, a robust approach was formulated for systems with fully decoupled dynamics but subject to coupled constraints. A similar vein of work was considered for systems with coupled dynamics via states in \cite{conte2013robust} and via inputs in \citeA{Walid2011}, and via both states and input in \citeA{grancharova2014approach}. In \citeA{zhang2014robust}, a robust MPC was also designed for a networked system with time delays. In these existing results, fixed terminal sets were used in the robust DMPC formulations.

Recently, \citeA{Conte2016} and \citeA{Darivianakis2020} consider the DMPC problem with the local terminal sets adapted online. In~\citeA{Conte2016}, time-varying local terminal sets are defined with a relaxation term from the Lyapunov function for decentralised systems from~\citeA{Jokic2009}. In~\citeA{Darivianakis2020}, adaptive local terminal sets are defined, where the parameters of these sets are chosen as decision variables to be determined in the same optimisation problem as DMPC. In both cases, the adaptation is motivated by a desire to avoid unnecessarily small terminal sets, which can adversely impact performance. However, the class of systems considered in this work does not explicitly consider disturbances in the system model. An open challenge is not only to implement a robust constraint tightening approach for distributed systems but also to properly define time-varying local terminal sets taking into account the effects of disturbances.


The main contribution of this work is to build on the work described in the previous two paragraphs to propose a robust DMPC for discrete-time linear systems subject to unknown-but-bounded disturbances. We formulate a robust DMPC that admits a separable structure along with appropriate design rules, in order to facilitate distributed optimisation techniques for implementation. Specifically,
\begin{itemize}
	\item We introduce \emph{robust adaptive local terminal sets}, whose sizes are determined online;
	\item We prove the closed-loop control system recursively feasible and input-to-state (ISS) stable;
	\item We present a synthesis method for a robust DMPC controller and develop design conditions for those local terminal sets.
\end{itemize}

The paper organisation begins after the problem statement in Section~\ref{section:problem statement}. The robust DMPC is formulated in Section~\ref{section:robust DMPC}. The  closed-loop performance analysis is discussed in Section~\ref{section:analysis}. The distributed synthesis method, the design conditions for terminal constraint, as well as summary of robust DMPC algorithm  are presented in Section~\ref{section:synthesis}. Implementation of the synthesis results and validation of the theoretical results are undertaken via simulation in Section~\ref{section:simulation}. Finally, the conclusion is drawn in Section~\ref{section:conlcusion}.

\paragraph*{Notation.}

We use~$ I $ to denote an identity matrix of appropriate dimension. For a matrix $ A $, we denote $ \mathrm{tr}(A) $ and $ \mathrm{rank}(A) $ as the trace and the rank of $ A $, $ A^{-1} $ and $ A^{\top} $ as the inverse and the transpose of $ A $, and $ A \succ 0 $ being the positive definiteness. For two matrices~$ A $ and~$ B $, we use $ \mathrm{diag}(A,B) $ to denote a block diagonal matrix. For a set of matrices~$ A_{j} $ with $ j \in N $,  we denote $ {\mathrm{col}}_{j \in \mathcal{N}}\{A_j\} := \left[ A_{j_1}^{\top} , A_{j_2}^{\top} ,\ldots, A_{j_N}^{\top}  \right]^{\top}  $, where $ j_{1} < j_{2} < \cdots < j_N $ are the (ordered) elements of~$ N $. Besides, we define the following sets $ \mathbb{S}^{n} := \left\lbrace X \in \mathbb{R}^{n \times n} \mid X = X^{\top} \right\rbrace  $, $ \mathbb{S}^{n}_{\succ 0} := \{ X \in \mathbb{R}^{n \times n} \mid  X = X^{\top}, X \succ 0 \} $ and $ \mathbb{S}^{n}_{\succeq 0} := \left\lbrace X \in \mathbb{R}^{n \times n} \mid  X = X^{\top}, X \succeq 0 \right\rbrace $. For a vector~$ z \in \mathbb{R}^{n} $ and a matrix~$ W \in \mathbb{S}^{n} $, we use~$ \left\| z \right\| $ to denote the 2-norm and the weighted 2-norm by $ W $, respectively. We use $ \mathrm{col}_{j \in \mathcal{N}} \left\lbrace z_j \right\rbrace  $ to denote the column vector with elements given by the vector $ z_j $, $ \forall j \in \mathcal{N} $. We use $ \mathrm{diag}(z) $ to denote a diagonal matrix with elements of its argument~$ z $.  For any two sets $ \mathcal{X} $ and $ \mathcal{Y} $, the Minkowski sum, Pontryagin difference and Cartesian product are denoted as $ \mathcal{X} \oplus \mathcal{Y} = \lbrace x+y : x\in \mathcal{X}, y\in \mathcal{Y} \rbrace $, $ \mathcal{X} \ominus \mathcal{Y} = \lbrace z: z+y \in \mathcal{X}, \forall y \in \mathcal{Y} \rbrace $, $ \mathcal{X} \times \mathcal{Y} = \lbrace (x,y): x \in \mathcal{X} \; \mathrm{and} \; y \in \mathcal{Y} \rbrace $, respectively. Besides, for sets $ \mathcal{X}_j $ with $ \forall j \in \mathcal{N}  $, we denote $ \bigoplus_{j \in \mathcal{N}} \mathcal{X}_j := \mathcal{X}_{j_1} \oplus \cdots \oplus \mathcal{X}_{j_N}  $ and $ \bigtimes_{j \in \mathcal{N}} \mathcal{X}_j := \mathcal{X}_{j_1} \times \cdots \times \mathcal{X}_{j_N}  $.

\section{Problem statement}\label{section:problem statement}

Let us consider a network of $M$ nodes (or called \emph{agents}). The agents exchange information according to a fixed graph defined by $\mathcal{G} := (\mathcal{M}, \mathcal{D})$, where the vertex set $\mathcal{M} := \left\lbrace 1,\ldots,M \right\rbrace $ indicates all the agents and the edge set $\mathcal{D} \subset \mathcal{M} \times \mathcal{M} $ specifies pairs of agents that can communicate. In general, the network admits the following linear time-invariant system dynamics:
\begin{equation}\label{eq: global linear system}
	x(k+1) = A x(k) + B u(k) + w(k),
\end{equation}
where $ x \in \mathbb{R}^{n} $, $ u \in \mathbb{R}^{m} $ denote the state vector and the control input vector and the discrete-time index $ k \in \mathbb{N} $. $ A \in \mathbb{R}^{n \times n} $ and $ B \in \mathbb{R}^{n \times m} $. For each agent $ i \in \mathcal{M} $, coupled dynamics via states is considered as follows:
\begin{equation}\label{eq:simplified subsystems}
	x_{i}(k+1) = A_{\mathcal{N}_{i}} x_{\mathcal{N}_{i}}(k) + B_{i} u_{i}(k) + w_{i}(k) ,\;\forall i \in \mathcal{M} ,
\end{equation}
where~$ x_{i} \in \mathbb{R}^{n_i} $, $ u_{i} \in \mathbb{R}^{m_i} $ denote the local state vector and the local control input vector, $ w_{i} \in \mathbb{R}^{n_i} $ denotes the local disturbance vector of the $ i $-th agent, respectively. $ A _{\mathcal{N}_{i}} := \mathrm{col}_{j \in \mathcal{N}_i}\{A_{ij}\} \in \mathbb{R}^{n_i \times n_{\mathcal{N}_i}}$, $ A_{ij} \in \mathbb{R}^{n_i \times n_j} $ and $ B_{i} \in \mathbb{R}^{n_i \times m_i} $. $ \mathcal{N}_{i} \subseteq \mathcal{M} $ is defined as the set that includes all the agents related to the agent $ i $ ($ i $ also included). Furthermore, local variables can be selected by using lifting matrices $ T_i \in \left\lbrace 0, 1 \right\rbrace^{n_{i} \times n}  $, $ L_i \in \left\lbrace 0, 1 \right\rbrace^{m_{i} \times m}  $, and $ T_{\mathcal{N}_i} \in \left\lbrace 0, 1 \right\rbrace^{n_{\mathcal{N}_i} \times n}  $ such that 
\begin{equation}
	x_i = T_i x , \; u_i = L_i u ,\; x_{\mathcal{N}_{i}}  = T_{\mathcal{N}_i}  x.
\end{equation}

\begin{assumption}\label{assump:controllability}
	For the global system~\eqref{eq: global linear system}, the pair $ (A,B) $ is controllable. The closed-loop system states $ x_i(k) $ can be measured at each time step $ k \in \mathbb{N} $.
\end{assumption}

\begin{assumption}
	The network graph is undirected, that is, any two neighbouring agents $ i \in \mathcal{N}_{j} $ and $ j \in \mathcal{N}_{i} $ can communicate with information exchange in a bidirectional way.
\end{assumption}

We are solving a finite horizon problem minimising $ \sum_{k=0}^{N} \ell (x(k), u(k)) $, $ N > 0 $ , where
\begin{equation}\label{eq:general control objective}
	\ell (x(k), u(k)) :=  \left\| x (k)\right\|_{Q}^{2} + \left\| u(k) \right\|_{R}^{2},
\end{equation}
with $ Q \in \mathbb{S}^{n}_{\succ 0} $ is a positive-definite weighting matrix, and $ R \in \mathbb{S}^{m}_{\succ 0} $ is a block-diagonal positive-definite weighting matrix. In addition, \eqref{eq:general control objective} can be rewritten in a separable structure
\begin{equation}\label{eq:quadratic cost function}
		\begin{split}
			\ell (x(k), u(k)) &= \sum_{i \in \mathcal{M}} \ell_i (x_{\mathcal{N}_i}(k),u_i(k)) \\
										&= \sum_{i \in \mathcal{M}} \big( \left\| x_{\mathcal{N}_i} (k)\right\|_{Q_{\mathcal{N}_i}}^{2} + \left\| u_i(k) \right\|_{R_{i}}^{2} \big)
		\end{split}
\end{equation}
where $ Q_{\mathcal{N}_i} = T_{\mathcal{N}_i} Q T_{\mathcal{N}_i}^{\top} \in \mathbb{S}^{n_{\mathcal{N}_i}}_{\succ 0} $ and $ R_{i} = L_{i} R L_{i}^{\top} \in \mathbb{S}^{m_{i}}_{\succ 0} $, $ \forall i \in \mathcal{M} $.

The system states and control inputs are constrained in convex sets
\begin{equation}\label{eq:closed-loop constraints}
	x(k) \in \mathcal{X} := \bigtimes_{i \in \mathcal{M}} \mathcal{X}_{i},\;
	u(k) \in \mathcal{U} := \bigtimes_{i \in \mathcal{M}}  \mathcal{U}_{i},
\end{equation}
for $ k \in \mathbb{N} $, where $ \mathcal{X}_i \subseteq \mathbb{R}^{n_i} $ and $ \mathcal{U}_i \subseteq \mathbb{R}^{m_i} $, $ \forall i \in \mathcal{M} $ are convex sets.

\begin{assumption}\label{assump:W boundedness}
	The disturbance vector $ w(k) $ is unknown but bounded by convex sets
	\begin{equation}
		w(k) \in \mathcal{W} := \bigtimes_{i \in \mathcal{M}}  \mathcal{W}_{i}, \; \forall k \in \mathbb{N}.
	\end{equation}
\end{assumption}

A nominal distributed model is now introduced as follows:
\begin{align}\label{eq:nominal simplified subsystems}
	\bar{x}_{i}(k+1) = A_{\mathcal{N}_{i}} \bar{x}_{\mathcal{N}_{i}}(k) + B_{i} \bar{u}_{i}(k), \;\forall i \in \mathcal{M}.
\end{align}

From the nominal system~\eqref{eq:nominal simplified subsystems}, the resulting global system can be formulated as
\begin{align}\label{eq:global nominal system}
	\bar{x}(k+1) = A \bar{x}(k) + B \bar{u}(k),
\end{align}
where $ \bar{x} = \mathrm{col}_{i \in \mathcal{M}} (\bar{x}_i) $ and $ \bar{u} = \mathrm{col}_{i \in \mathcal{M}} (\bar{u}_i) $.

For the system~\eqref{eq: global linear system}, centralised robust MPC formulation can be formulated~\cite{Mayne2005}
\begin{subequations}\label{problem:RMPC}
	\begin{align}
	&\underset{\bar{u}(0),\ldots,\bar{u}(N-1)} {\mathrm{minimise}}\;\;  V_{f}(\bar{x}(N)) +\sum_{t=0}^{N-1} \ell (\bar{x}(t),\bar{u}(t)),
	\intertext{subject to}
	& \bar{x}(t+1) = A\bar{x}(t) + B \bar{u} (t),\\
	& \bar{x}(t) \in \bar{\mathcal{X}}, \label{eq:RMPC constraint-xbar}\\
	& \bar{u}(t) \in \bar{\mathcal{U}}, \label{eq:RMPC constraint-ubar}\\
	& \bar{x}(N) \in \Omega_{f},\label{eq:RMPC terminal constraint}\\
	& \bar{x}(0) = x(k),\label{eq:RMPC initialisation}
	\end{align}
\end{subequations}
where $ \bar{\mathcal{X}} $ and $ \bar{\mathcal{U}} $ are tightened sets for states and inputs. $ V_{f}(\bar{x}(N)) $ and $ \Omega_{f} $ are terminal cost function and terminal set to guarantee the closed-loop convergence.

To formulate a distributed solution for~\eqref{problem:RMPC}, the following problems must be considered:

\begin{problem}\label{OP:1}
	How can the constraints~$ \bar{\mathcal{X}} $ and~$ \bar{\mathcal{U}} $ be tightened for distributed systems with disturbances?
\end{problem} 

\begin{problem}\label{OP:2}
	How can the terminal cost~$ V_{f} $ and the terminal set~$ \Omega_{f} $ be defined in a distributed way?
\end{problem}

\section{Robust DMPC based on set-membership approach}\label{section:robust DMPC}

In this section, we formulate the robust DMPC considering a finite prediction horizon $ N > 0 $. We first propose a constraint tightening approach based on set-membership approach. Besides, we also use another auxiliary terminal gain to find separable terminal cost function and local terminal sets.

\subsection{Robust constraint tightening approach}\label{subsec:constraint tightening}

Since system states are coupled in the system dynamics~\eqref{eq:simplified subsystems}, the effect of disturbance $ w_i (k) $ for agent $ i $ will be propagated into other neighbours. To solve Problem~\ref{OP:1}, we propose a robust constraint tightening approach to tighten constraints on states and inputs iteratively along the MPC prediction horizon~$ N $ in terms of the global systems~\eqref{eq: global linear system} and~\eqref{eq:global nominal system}, and then project the tightened constraints into each agent.

For the system~\eqref{eq: global linear system}, there exists a state feedback controller
\begin{equation}\label{eq: K}
	u = \kappa(x) :=  K x, 
\end{equation}
where $ K \in \mathbb{R}^{m \times n} $ is a stabilising state feedback gain.

By comparing the systems~\eqref{eq: global linear system} and~\eqref{eq:global nominal system}, let us define the error
$ e := x  - \bar{x} $, and therefore the error dynamics can be formulated with the control law in~\eqref{eq: K}  as
\begin{align}\label{eq:error dynamics}
	e(k+1) = A_{K} e (k) + w(k), \; \forall k \in \mathbb{N},
\end{align}
where $ A_{K} := A+ BK $.

Set $ \bar{x}(0) = x(k) $. Then, we can derive $ e(0) = 0 $, $ e(1) = A_{K} e(0) + w(1) = w(1) $, and for $ t \geq 1 $, $	e(t+1) =  A_{K} e (t) + w(t) $.

\begin{remark}
	In order to clarify the different time steps, we use $ t \in \left\lbrace 0,1,\ldots, N \right\rbrace $ to denote the MPC prediction steps while $ k \in \mathbb{N} $ to denote the closed-loop simulation steps.
\end{remark}

The closed-loop states and inputs are constrained in the convex sets, that is, $x \in \mathcal{X}$ and $u \in \mathcal{U}$. Based on the above analysis, from \eqref{eq:error dynamics} with $ w(k) \in \mathcal{W} $ under Assumption~\ref{assump:W boundedness}, the constraints on nominal states and inputs can be tightened as follows:
\begin{subequations}\label{eq:tightened constraints}
	\begin{align} 
		\bar{\mathcal{X}}(t) &:=\mathcal{X} \ominus \mathcal{R}(t), \; t=1,\ldots, N-1, \\
		\bar{\mathcal{U}}(t) &:=\mathcal{U} \ominus K \mathcal{R}(t), \; t=1,\ldots, N-1
	\end{align}
\end{subequations}
with $\bar{\mathcal{X}}(0) :=\mathcal{X} $ and $\bar{\mathcal{U}}(0) :=\mathcal{U}$, where the set $ \mathcal{R}(t) $ is defined as 
\begin{align}\label{eq:set definition R-Q}
	\mathcal{R}(t) := \bigoplus_{j=0}^{t-1}  A_{K}^{j} \mathcal{W}, \; t \geq 1.
\end{align}

From the definition of $\mathcal{R}(t)$ in \eqref{eq:set definition R-Q}, it holds
\begin{equation}\label{eq:R property}
	\mathcal{R}(t+1) \ominus A_{K}^{t}  \mathcal{W} = \mathcal{R}(t),\; t \geq 1, \; \forall i \in \mathcal{M}.
\end{equation}

\begin{remark}
	Note it is possible to generalise the problem formulation slightly to consider bounded measurement error at time step $ k $, so that $ e(0) \ne 0 $. This will incur further refinement of $ \mathcal{R}(t) $ in~\eqref{eq:set definition R-Q}, but all subsequent steps in the following theoretical results are maintained.
\end{remark}

 \begin{assumption}\label{assump:K}
	There exists a $K$ guaranteeing that the matrix $ A_{K}$ is Schur stable, and  for the chosen $K$ the sets $ \bar{\mathcal{X}}(N) $ and $ \bar{\mathcal{U}}(N) $  from~\eqref{eq:tightened constraints} are non-empty. 	
 \end{assumption}

A synthesis method is proposed in Section \ref{section:synthesis} to find a $K$ if Assumption \ref{assump:K} holds.

From the tightened constraints in~\eqref{eq:tightened constraints}, we can find the underlying constraints for each agent $ i $ by the projection with the lifting matrices.
\begin{subequations}\label{eq:tightened local constraints}
	\begin{align}
		\bar{\mathcal{X}}_{\mathcal{N}_i}(t) &= T_{\mathcal{N}_i} \bar{\mathcal{X}}(t),\; \forall  i \in \mathcal{M},\\
		\bar{\mathcal{U}}_i(t) &= L_i \bar{\mathcal{U}}(t), \; \forall  i \in \mathcal{M}.
	\end{align}
\end{subequations}

\begin{remark}\label{remark:Kni}
	The state and input constraints can be also tightened locally for each agent if there exist local stabilising feedback control laws
	\begin{equation*}
		u_i = \kappa_{i} \left( x_{\mathcal{N}_i} \right) := K_{\mathcal{N}_i} x_{\mathcal{N}_i},\; \forall i \in \mathcal{M}, 
	\end{equation*}
	where $ K_{\mathcal{N}_i}  \in  \mathbb{R}^{m_i \times n_{\mathcal{N}_i}} $ is a local control gain. Since the local disturbance vector is bounded in a separable set $ w_i \in \mathcal{W}_i $, we can also tighten the constraints locally to find $ \bar{\mathcal{X}}_i $ and $ \bar{\mathcal{U}}_i $, $ \forall i \in \mathcal{M} $. If all stabilising $K_{\mathcal{N}_i}$ are found, the global feedback gain $K$ may be determined using local gains and the lifting matrices, i.e.
	\begin{equation*}
		K = \sum_{i \in \mathcal{M}} L_i^{\top} K_{\mathcal{N}_i} T_{\mathcal{N}_i}.
	\end{equation*}
\end{remark}

\subsection{Separable terminal cost}

We now turn our attention to Problem~\ref{OP:2}. To set up a distributed optimisation problem, since the stage cost function and constraints can be set in a distributed way as in~\eqref{eq:general control objective} and~\eqref{eq:tightened local constraints}, we formulate the separable terminal cost function as
\begin{equation}\label{eq:global terminal cost function}
	V_f (x) = \sum_{i \in \mathcal{M}} V_{f_i}(x_i) = \sum_{i \in \mathcal{M}} x_i^{\top} P_{f_i} x_i,
\end{equation}
with $ P_{f_i} \in \mathbb{S}^{n_i}_{\succ 0} $, and terminal control law can be defined as
\begin{equation}\label{eq:terminal control gain}
	u_i = \kappa_{f_i}(x_{\mathcal{N}_i}) := K_{f_i} x_{\mathcal{N}_i}, \; \forall i \in \mathcal{M},
\end{equation}
where $ K_{f_i} \in \mathbb{R}^{m_i \times n_{\mathcal{N}_i}} $ is a terminal control gain.

The following lemma indicates the local terminal cost may be increasing with a relaxation term but the combination of the relaxation terms across all agents leads to a strictly decreasing global terminal cost.

\begin{lemma}[\cite{Jokic2009}]\label{lemma:separable terminal cost}
	If there exists the functions $ V_{f_i}(x_i) $, $\gamma_i (x_{\mathcal{N}_i}) $ and $ \ell_i (x_{\mathcal{N}_i},K_{f_i} x_{\mathcal{N}_i}) $, as well as $ \mathcal{K}_{\infty} $ functions $ \beta_{1_i} $, $ \beta_{2_i} $ and $ \beta_{3_i} $, $ \forall i \in \mathcal{M} $ such that
	\begin{subequations}\label{eq:conditions for separable terminal cost}
		\begin{align}
			& \beta_{1_i} (\left\| x_i \right\| ) \leq V_{f_i}(x_i) \leq \beta_{2_i} (\left\| x_i \right\| ),\\
			& \beta_{3_i} (\left\| x_{\mathcal{N}_i} \right\| ) \leq \ell_i (x_{\mathcal{N}_i},K_{f_i} x_{\mathcal{N}_i}),\\
			& V_{f_i}(A_{\mathcal{N}_i} x_{\mathcal{N}_i} + B_i \kappa_{f_i}(x_{\mathcal{N}_i})) - V_{f_i}(x_i) \nonumber\\
			& \qquad \leq -\ell_i (x_{\mathcal{N}_i},\kappa_{f_i}(x_{\mathcal{N}_i})) + \gamma_i (x_{\mathcal{N}_i}) ,\label{eq:negativity of local terminal cost}\\
			& \sum_{i \in \mathcal{M}} \gamma_i (x_{\mathcal{N}_i}) \leq 0,\; i \in \mathcal{M},\label{eq:total gamma}
		\end{align}
	\end{subequations}
	then $ V_f(x) = \sum_{i \in \mathcal{M}} V_{f_i} (x_i)  $ defined in~\eqref{eq:global terminal cost function} is a Lyapunov function for the system~\eqref{eq:nominal simplified subsystems} with $ u_i = \kappa_{f_i}(x_{\mathcal{N}_i})$ defined in~\eqref{eq:terminal control gain}, $ \forall i \in \mathcal{M} $.
\end{lemma}

\subsection{Robust adaptive local terminal sets}

We next discuss how to choose local terminal sets for each agent. In general, let us define the local terminal sets as
\begin{equation}\label{eq:local terminal set}
	\Omega_{f_i} (\alpha_i) := \left\lbrace x_i \in \mathbb{R}^{n_i} : x_i^{\top} F_i x_i \leq  \alpha_i  \right\rbrace ,\; \forall i \in \mathcal{M},
\end{equation}
where $ F_{i} \in \mathbb{S}^{n_i}_{\succ 0} $ and a scalar $ \alpha_i > 0$ determine the size of local terminal sets. For each agent $ i $, since $ V_{f_i} $ satisfies the conditions~\eqref{eq:conditions for separable terminal cost}, it can be seen that the local terminal cost might increase. Therefore, the corresponding local terminal set should be also adaptive along the terminal cost increasing.

Considering a prediction horizon of $ N $, the constraints on states and inputs are tightened iteratively as in~\eqref{eq:tightened local constraints}. We now define the local terminal sets $ \Omega_{f_i} (\alpha_i) $ for the robust DMPC controller with updating parameters.

\begin{definition}[Robust adaptive local terminal sets]\label{definition:terminal sets}
	For each agent $ i \in \mathcal{M} $, the set $ \Omega_{f_i} (\alpha_i) $ is said to be a robust adaptive local terminal set if there exists a matrix $ F_{i} \in \mathbb{S}^{n_i}_{\succ 0} $ and a scalar $ \alpha_i > 0$ such that
	\begin{equation}\label{eq:robust adaptive termianl set}
		A_{K_{f_i}} \left\lbrace \bar{x}_{\mathcal{N}_i}  \oplus  \bar{\mathcal{E}}_{\mathcal{N}_i} (N-1)  \right\rbrace \in  \Omega_{f_i} (\alpha_i),
	\end{equation}
	where $ \bar{\mathcal{E}}_{\mathcal{N}_i}  (N-1) := T_{\mathcal{N}_i}  A_{K}^{N-1} \mathcal{W} $ and $ A_{K_{f_i}} := A_{\mathcal{N}_i} + B_i K_{f_i} $.
\end{definition}

\begin{remark}
	For the selection of $ F_i $ and $ \alpha_i $, we can set $ F_i = P_{f_i} $ as widely used in~\shortciteA{Maestre2011,Conte2016,Darivianakis2020} and update the scalar $ \alpha_i $ online.
\end{remark}

\subsection{Robust DMPC optimisation problem}

Based on the discussions above, we now formulate the optimisation problem of the robust DMPC in the following.
\begin{subequations}\label{problem:RDMPC}
	\begin{align}
	&\underset{\substack{\bar{u}_i(0),\ldots,\bar{u}_i(N-1)\\\alpha_1,\ldots,\alpha_M}} {\mathrm{minimise}}\;\;  \sum_{i=1}^{M} \Big( V_{f_i}(\bar{x}_i(N)) +\sum_{t=0}^{N-1}\ell_i (\bar{x}_{\mathcal{N}_i}(t),\bar{u}_i(t)) \Big) ,\label{eq:MPC cost function}
	\intertext{subject to}
		& \bar{x}_{i}(t+1) = A_{\mathcal{N}_{i}} \bar{x}_{\mathcal{N}_{i}}(t) + B_{i} \bar{u}_{i} (t),\label{eq:MPC prediction model}\\
		& \bar{x}_{\mathcal{N}_i}(t) \in \bar{\mathcal{X}}_{\mathcal{N}_i}(t),\label{eq:MPC constraint-xbar}\\
		& \bar{u}_i(t) \in \bar{\mathcal{U}}_i(t) ,\label{eq:MPC constraint-ubar}\\
		& \bar{x}_i(N) \in \Omega_{f_i} (\alpha_i),\label{eq:MPC terminal constraint}\\
		& \bar{x}_i(0) = x_i(k).\label{eq:MPC initialisation}
	\end{align}
\end{subequations}

The optimisation problem~\eqref{problem:RDMPC} can be implemented by means of alternative distribution optimisation techniques, such as dual decomposition or ADMM~\shortciteA{Farokhi2014,Boyd2011}. We use the superscript $ * $ to denote the variables related to the optimal solutions of~\eqref{problem:RDMPC}. For instance, let us denote the feasible solutions of~\eqref{problem:RDMPC} at time step $ k \in \mathbb{N} $ as follows:
\begin{subequations}\label{eq:RDMPC feasible solutions}
	\begin{align}
		& \bar{x}_{\mathcal{N}_i}^{*}(0;x_{i}(k)),\ldots,\bar{x}_{\mathcal{N}_i}^{*}(N;x_i(k)),\label{eq:feasible x of DMPC}\\
		& \bar{u}_{i}^{*}(0;x_{i}(k)), \ldots, \bar{u}_{i}^{*}(N-1;x_{i}(k)),\label{eq:feasible u of DMPC}
	\end{align}
\end{subequations}
and $ \alpha_i^* $, $ \forall i \in \mathcal{M} $. Therefore, by proceeding with the receding-horizon strategy, the optimal MPC law can be chosen for the closed-loop system at the time step $ k $ as
\begin{equation}\label{eq:RDMPC control action}
	\kappa_{N} (x_i(k)) := \bar{u}^{*}_{i} (0;x_i(k)),\; \forall i \in \mathcal{M} .
\end{equation}

\section{Properties of the closed-loop system}\label{section:analysis}

We now analyse the properties of the closed-loop system~\eqref{eq:simplified subsystems} operated by the robust DMPC controller~\eqref{problem:RDMPC}. 



Since the prediction model~\eqref{eq:MPC prediction model} does not contain disturbances, there exists a mismatch between the predicted states and the closed-loop states. Based on the constraint tightening approach in Section~\ref{subsec:constraint tightening}, we used an auxiliary control gain $ K$ to attenuate the effect of this mismatch in closed-loop. Similar to the robust tube-based technique originally proposed in~\cite{Mayne2005}, the optimal control action at time step $ k $ can be chosen as 
\begin{equation}\label{eq:tube-based control action}
	u_i(k) := \bar{u}_{i}^{*}(0;x_{i}(k)) + L_i K(x(k) - \bar{x}^{*}(0;x(k))).
\end{equation}

\begin{remark}
	If the local feedback control gains $ K_{\mathcal{N}_i} $ are chosen for constraint tightening in Section~\ref{subsec:constraint tightening}, then the control action~\eqref{eq:tube-based control action} can be adapted to be
	\begin{equation*}
		u_i(k) := \bar{u}_{i}^{*}(0;x_{i}(k)) + K_{\mathcal{N}_i} (x_{\mathcal{N}_i}(k) - \bar{x}_{\mathcal{N}_i}^{*}(0;x_{i}(k))),
	\end{equation*}
	where the mismatch of local states for Agent $ i $ is attenuated by local feedback gain $ K_{\mathcal{N}_i} $.
\end{remark}

With the proposed robust DMPC controller, recursive feasibility of the closed-loop system is summarised.

\begin{theorem}[Recursive feasibility]\label{theorem:recursive feasibility}
	Consider that Assumptions~\ref{assump:controllability}-\ref{assump:W boundedness} and the conditions of Lemma~\ref{lemma:separable terminal cost} hold. For any feasible initial condition $ x_i (0) $, $ \forall i \in \mathcal{M} $, the closed-loop system~\eqref{eq: global linear system} with~\eqref{problem:RDMPC} is recursively feasible.
\end{theorem}

\begin{pf}	
	See Appendix~\ref{appendix:recursive feasibility}.
\end{pf}

Since the closed-loop system is recursively feasible, we next consider the closed-loop stability.

\begin{theorem}[ISS stability]\label{theorem:ISS stability}
	Consider that Assumptions~\ref{assump:controllability}-\ref{assump:W boundedness} and the conditions of Lemma~\ref{lemma:separable terminal cost} hold. For any feasible initial condition $ x_i (0) $, $ \forall i \in \mathcal{M} $, the closed-loop system~\eqref{eq: global linear system} with~\eqref{eq:quadratic cost function} and~\eqref{problem:RDMPC} is ISS stable.
\end{theorem}

\begin{pf}
	See Appendix~\ref{appendix:ISS stability}.
\end{pf}

\section{Synthesis and design for robust DMPC}\label{section:synthesis}

In this section, we propose a synthesis method to design feedback control gains and robust adaptive local terminal sets needed for implementation of the theoretical results underpinning the robust DMPC algorithm that were introduced in Section~\ref{section:robust DMPC}.

\subsection{Synthesis of local feedback gains}

We first present a synthesis method to find a global feedback gain $ K $ as well as local feedback gains $ K_{\mathcal{N}_i} $, $ \forall i \in \mathcal{M} $.

\subsubsection{Synthesis of $ K $}

Consider the convex sets in the polytopic forms:
\begin{subequations}\label{eq:convex sets X-U}
	\begin{align}
	\mathcal{X} &:= \left\lbrace x \in \mathbb{R}^{n} :  a_{i}^{\top} x \leq d_i, i=1,\ldots,n_{r} \right\rbrace,\label{eq:convex sets X}\\
	\mathcal{U} &:= \left\lbrace u \in \mathbb{R}^{m} : h_{j}^{\top} u \leq g_j, j=1,\ldots,m_{r}  \right\rbrace, \label{eq:convex sets U}
	\end{align}
\end{subequations}
where $ a_i \in \mathbb{R}^{n} $, $ h_j \in \mathbb{R}^{m} $, $ d_i \in \mathbb{R} $, $ g_{j} \in \mathbb{R} $ with non-zero elements, and $ n_{r} $ and $ m_{r} $ are the number of linear constraints of $ \mathcal{X} $ and $ \mathcal{U} $.

We also consider the disturbance set.
\begin{equation}\label{eq:convex set W}
\begin{split}
\mathcal{W} := &\left\lbrace w \in \mathbb{R}^{n} : \left| w \right| \leq v \right\rbrace\\ 
:=&\left\lbrace w \in \mathbb{R}^{n} : w^{\top} W w \leq 1 \right\rbrace,
\end{split}
\end{equation}
where $ v \in \mathbb{R}^{n} $ with assuming non-zero elements and a diagonal matrix $ W \in \mathbb{S}^{n}_{\succ 0} $. Besides, $ \mathcal{W} = 	\mathcal{W}_1 \times \cdots \times 	\mathcal{W}_{n} $.

In general, the synthesis objectives for $ K $ are concluded as follows:
\begin{itemize}
	\item For the system~\eqref{eq: global linear system} with~\eqref{eq: K}, there exists a matrix $ P \in \mathbb{S}^{n}_{\succ 0} $ such that the set
	\begin{equation}\label{eq:global RPI set}
		\mathcal{Z} := \left\lbrace x \in \mathbb{R}^{n}: x^{\top} P x \leq 1 \right\rbrace ,
	\end{equation}
	is a minimum RPI set, $ \forall k \in \mathbb{N} $, $ \forall w \in \mathcal{W} $. From the definition of $ \mathcal{R}(t) $, it holds $  \mathcal{R}(t)  \subseteq \mathcal{Z}$, $ \forall t \in \mathbb{N} $.
	\item The tightened constraint sets $ \bar{\mathcal{X}}(N) $ and $ \bar{\mathcal{U}}(N) $ are non-empty.
\end{itemize}

Based on these objectives, the following result provides a condition to find the RPI set $ \mathcal{Z} $.

\begin{theorem}\label{theorem:RPI set}
	Given the set $ W $ defined in~\eqref{eq:convex set W}. If there exist matrices $ S \in \mathbb{S}^{n}_{\succ 0} $, $ Y \in \mathbb{R}^{m \times n} $, and two scalars $ \tau_1 \geq 0 $, $ \tau_2 \geq 0 $  such that 
	\begin{subequations}\label{eq:synthesis K RPI}
		\begin{align}
			& \begin{bmatrix}
				\tau_2 W && I && 0 \\
				\star && S && AS+BY\\
				\star && \star && \frac{1}{\tau_1} S
			\end{bmatrix} \succeq 0, \label{eq:synthesis K-1}\\
			&  \tau_1 + \tau_2 \leq 1,\label{eq:synthesis K-2}
		\end{align}
	\end{subequations}
	then $ \mathcal{Z} $ is an RPI set, that is, $ x(k+1) \in \mathcal{Z} $, for any $ x(k) \in \mathcal{Z} $, $ \forall w(k) \in \mathcal{W}$, $ \forall k \in \mathbb{N} $.  Moreover, $ P = S^{-1} $ and $ K=Y S^{-1} $.
\end{theorem}

\begin{pf}
	See Appendix~\ref{appendix:RPI ellipsoid}. 
\end{pf}

Another objective is to make sure $ \bar{\mathcal{X}}(N) $ and $ \bar{\mathcal{U}}(N) $ are non-empty, which can be satisfied if $ \mathcal{X} \ominus \mathcal{Z} $ and $ \mathcal{U} \ominus K \mathcal{Z} $ are non-empty due to $ R(t) \subseteq \mathcal{Z} $, $ \forall t \in \mathbb{N} $. We give the corresponding conditions in the following theorem.

\begin{theorem}\label{theorem:K nonempty sets}
	Given the convex sets in~\eqref{eq:convex sets X-U}-\eqref{eq:convex set W}. For the RPI set $ \mathcal{Z} $ in~\eqref{eq:global RPI set}, if there exist matrices $ S \in \mathbb{S}^{n}_{\succ 0} $, $ Y \in \mathbb{R}^{m \times n} $ such that
	\begin{subequations}\label{eq:synthesis K nonempty sets}
		\begin{align}
			&\begin{bmatrix}
				d_i^2 && a_i^{\top} S\\
				\star && S
			\end{bmatrix} \succeq 0, \; i =1,\ldots, n_r,\label{eq:synthesis K-3}\\
			& \begin{bmatrix}
				g_j^2 && h_j^{\top} Y \\
				\star && S
			\end{bmatrix} \succeq 0, \; j =1,\ldots, m_r,\label{eq:synthesis K-4}
		\end{align}
	\end{subequations}
	then the sets $ \mathcal{X} \ominus \mathcal{Z} $ and $ \mathcal{U} \ominus K \mathcal{Z} $ are non-empty. Moreover, $ P = S^{-1} $ and $ K=Y S^{-1} $.
\end{theorem}

\begin{pf}
	See Appendix~\ref{appendix:K nonempty sets}. 
\end{pf}


As a result, the auxiliary control gain $ K $ can be synthesised via offline solving the following optimisation with the objective of finding a minimum RPI set $ \mathcal{Z} $.
\begin{equation}\label{problem:K synthesis}
	\underset{S,G,Y,\tau_2,\mu} {\mathrm{minimise}}\;\;  \mathrm{trace}(S) ,
\end{equation}
subject to~\eqref{eq:synthesis K RPI}-\eqref{eq:synthesis K nonempty sets}, for given $ \tau_1 >0 $.

\begin{remark}
	If we choose a structured $ P $ for the Lyapunov candidate function, i.e.
	\begin{align*}
	V(x) &= \sum_{i \in \mathcal{M}} x_{\mathcal{N}_i}^{\top} P_{\mathcal{N}_i} x_{\mathcal{N}_i}
	= \sum_{i \in \mathcal{M}} x^{\top}  \bar{P}_i x,
	\end{align*}
	where $ P_{\mathcal{N}_i} \in \mathbb{S}^{n_i}_{\succ 0} $ and $ \bar{P}_i = T_{\mathcal{N}_i}^{\top} P_{\mathcal{N}_i}  T_{\mathcal{N}_i}  $, then we may find an RPI set for $ x_{\mathcal{N}_i} $ 
	\begin{align*}
		 \mathcal{Z}_{\mathcal{N}_i} := \left\lbrace x_{\mathcal{N}_i} \in \mathbb{R}^{n_{\mathcal{N}_i}}: x_{\mathcal{N}_i}^{\top} P_{\mathcal{N}_i} x_{\mathcal{N}_i} \leq \varphi_i \right\rbrace,  \;\forall i \in \mathcal{M},
	\end{align*}
	where $ P_{\mathcal{N}_i} \in \mathbb{S}^{n_{\mathcal{N}_i}}_{\succ 0} $, $ \varphi_i \geq 0 $ and $ \sum_{i \in \mathcal{M}}  \varphi_i \leq 1 $. The synthesis condition can be found in~\cite[(21)-(25)]{conte2013robust}.
\end{remark}

\subsubsection{Synthesis of $ K_{\mathcal{N}_i} $}

For a non-empty disturbance set, it may become more challenging to apply a centralised constraint tightening approach as the system order increases. Instead, a distributed robust constraint tightening approach can be implemented with local feedback control gains $ K_{\mathcal{N}_i} $, $ \forall i \in \mathcal{M} $ so that less conservative tightening may be applied. We next present the synthesis conditions for finding these local feedback control gains.

\begin{corollary}\label{corollary:distributed RPI set}
	For each agent $ i \in \mathcal{M} $, if there exist matrices $ S_i \in \mathbb{S}^{n_i}_{\succ 0} $ with $ S_{ij} = T_{\mathcal{N}_i} T_j^{\top} S_j T_j T_{\mathcal{N}_i}^{\top} $, $ G_i \in \mathbb{R}^{n_{\mathcal{N}_i} \times n_{\mathcal{N}_i}} $, $ Y_i \in \mathbb{R}^{m_i \times n_{\mathcal{N}_i}} $, and two scalars $ \bar{\tau}_i \geq 0 $, $ \bar{\tau}_{ij} \geq 0 $, $ \forall j \in \mathcal{N}_i $ such that 
	\begin{subequations}\label{eq:synthesis K_Ni RPI}
		\begin{align}
		& \begin{bmatrix}
		\bar{\tau}_i W_i && I && 0 \\
		\star && S_i && A_{\mathcal{N}_i} G_i + B_i Y_i\\
		\star && \star && G_i+G_i^{\top} -  \displaystyle \sum_{j \in \mathcal{N}_i}\frac{S_{ij}}{\bar{\tau}_{ij}} 
		\end{bmatrix} \succeq 0, \label{eq:synthesis Kni-1}\\
		&  \bar{\tau}_i + \displaystyle \sum_{j \in \mathcal{N}_i} \bar{\tau}_{ij} \leq 1,\label{eq:synthesis Kni-2}
		\end{align}
	\end{subequations}
	then $ \mathcal{Z}_i = \left\lbrace x_i \in \mathbb{R}^{n_i}: x_{i}^{\top} P_i x_{i} \leq 1 \right\rbrace  $ is an RPI set, that is, $ x_i(k+1) \in \mathcal{Z}_i $, $ \forall x_i(k) \in \mathcal{Z}_i $, $ \forall w_i(k) \in \mathcal{W}_i $, $ \forall k \in \mathbb{N} $. Moreover, $ P_i = S_i^{-1} $ and $ K_{\mathcal{N}_i} = Y_i G_i^{-1} $.
\end{corollary}

\begin{pf}
	See Appendix~\ref{appendix:proof-corollary 1}. 
\end{pf}

Deriving from~\eqref{eq:convex sets X-U}, the local constraints on states and inputs are considered as follows:
\begin{subequations}\label{eq:convex sets Xi-Ui}
	\begin{align}
	\mathcal{X}_i &:= \left\lbrace x_i \in \mathbb{R}^{n_i} :  a_{il}^{\top} x_i \leq d_{il}, l=1,\ldots,n_{r_i} \right\rbrace,\label{eq:convex sets Xi}\\
	\mathcal{U}_i &:= \left\lbrace u_i \in \mathbb{R}^{m_i} : h_{ip}^{\top} u_i \leq g_{ip}, p=1,\ldots,m_{r_i}  \right\rbrace, \label{eq:convex sets Ui}
	\end{align}
\end{subequations}
and the local disturbance set
\begin{equation}
	\begin{split}
		\mathcal{W}_i := & \left\lbrace w_i \in \mathbb{R}^{n_i} : \left| w_i \right| \leq v_i \right\rbrace\\
		=&\left\lbrace w_i \in \mathbb{R}^{n_i} : w_i^{\top} W_i w_i \leq 1 \right\rbrace .
	\end{split}
\end{equation}

\begin{corollary}\label{corollary:Kni nonempty sets}
	For each agent $ i \in \mathcal{M} $, if there exist matrices $ S_i \in \mathbb{S}^{n_i}_{\succ 0} $ with $ S_{ij} = T_{\mathcal{N}_i} T_j^{\top} S_j T_j T_{\mathcal{N}_i}^{\top} $, $ G_i \in \mathbb{R}^{n_{\mathcal{N}_i} \times n_{\mathcal{N}_i}} $, $ Y_i \in \mathbb{R}^{m_i \times n_{\mathcal{N}_i}} $, and scalars $ \tilde{\tau}_{ijp} \geq 0 $, $ \forall j \in \mathcal{N}_i $, $ p =1,\ldots, m_{r_i} $ such that
	\begin{align}\label{eq:synthesis Kni-3}
		\begin{bmatrix}
		d_{il}^2 && a_{il}^{\top} S_i\\
		\star && S_i
		\end{bmatrix} \succeq 0, \; l =1,\ldots, n_{ri},
	\end{align}
	and
	\begin{subequations}\label{eq:synthesis Kni nonempty sets}
		\begin{align}
			& \begin{bmatrix}
			g_{ip} && h_{ip}^{\top} Y_i \\
			\star &&  G_i+G_i^{\top} -  \displaystyle \sum_{j \in \mathcal{N}_i}\frac{S_{ij}}{\tilde{\tau}_{ijp}} 
			\end{bmatrix} \succeq 0, \; p =1,\ldots, m_{r_i},\label{eq:synthesis Kni-4}\\
			& \displaystyle \sum_{j \in \mathcal{N}_i} \tilde{\tau}_{ijp} \leq g_{ip},\label{eq:synthesis Kni-5}
		\end{align}
	\end{subequations}
	then the tightened constraint sets on states and inputs are non-empty. Moreover, $ P_i = S_i^{-1} $ and $ K_{\mathcal{N}_i}=Y_i G_i^{-1} $.
\end{corollary}

\begin{pf}
	The proof follows directly from the proofs of Theorem~\ref{theorem:K nonempty sets} and Corollary~\ref{corollary:distributed RPI set}. \qed
\end{pf}

\subsection{Synthesis of terminal gains and costs}

Considering the terminal cost function defined in~\eqref{eq:global terminal cost function}, for each agent $ i \in \mathcal{M} $, we have $ V_{f_i}(x_i) = x_{i}^{\top} P_{f_i} x_{i} $ with $ P_{f_i} \in \mathbb{S}^{n_i}_{\succ 0} $. We also consider the relaxation function as $ \gamma_i(x_{\mathcal{N}_i}) = x_{\mathcal{N}_i}^{\top} \varGamma_i x_{\mathcal{N}_i} $ with $  \varGamma_i \in \mathbb{S}^{n_{\mathcal{N}_i}}_{\succeq 0}$. The terminal gain $K_{f_i}$ and the matrix $P_{f_i}$ for $ i \in \mathcal{M} $ can be obtained by using \cite[Lemma 10]{Conte2016}.

\subsection{Conditions for robust local adaptive terminal sets}

To implement the optimisation problem~\eqref{problem:RDMPC}, we need conditions for the terminal constraint~\eqref{eq:MPC terminal constraint}. Based on Definition~\ref{definition:terminal sets}, we give the condition for robust adaptive local terminal sets in the following theorem.

\begin{theorem}\label{theorem:Omega_fi}
	Given the system~\eqref{eq:simplified subsystems} with $ K $, $ F_i $ and $ K_{f_i} $, $ \forall i \in \mathcal{M} $. For each agent $ i \in \mathcal{M} $, if there exist scalars $ \alpha_i $,  and $ \sigma_i \geq 0 $, $ \sigma_{ij} \geq 0 $, $ \forall j \in \mathcal{N}_i $ such that
	\begin{subequations}\label{eq:Omega_fi condition}
		\begin{align}
			& \begin{bmatrix}
				\displaystyle\sum_{j \in \mathcal{N}_i} \sigma_{ij} F_{ij} && ( \alpha_{\mathcal{N}_i}^{\frac{1}{2}} )^{\top} A_{K_{f_i}}^{\top} &&  0 \\
				\star && \alpha_{i}^{\frac{1}{2}} F_{i}^{-1} && A_{K_{f_i}}\\
				\star && \star && \sigma_i E_{\mathcal{N}_i}
			\end{bmatrix} \succeq 0\label{eq:Omega_fi condition-1}\\
			& \sigma_i + \sum_{j \in \mathcal{N}_i} \sigma_{ij} \leq \alpha_{i}^{\frac{1}{2}},\label{eq:Omega_fi condition-2}
		\end{align}
	\end{subequations}
	where $ \alpha_{\mathcal{N}_i} = T_{\mathcal{N}_i} \alpha T_{\mathcal{N}_i}^{\top} $, $ \alpha = \mathrm{diag}(\alpha_1 I_{n_1},\ldots, \alpha_M I_{n_M}) $, $ F_{ij} = T_{\mathcal{N}_i} T_i^{\top} F_{i} T_i T_{\mathcal{N}_i}^{\top} $, and $ E_{\mathcal{N}_i} = T_{\mathcal{N}_i} A_{K}^{N-1} W(T_{\mathcal{N}_i} A_{K}^{N-1})^{\top}  $, then the condition~\eqref{eq:robust adaptive termianl set} is satisfied.
\end{theorem}

\begin{pf}
	See Appendix~\ref{appendix:proof-Omega_fi}. 
\end{pf}

Let $\Omega_{f_{\mathcal{N}_i}} = \bigtimes_{j \in \mathcal{N}_i} \Omega_{f_j} (\alpha_j) $, $\forall i \in \mathcal{M}$. Any state $x_{f_{\mathcal{N}_i}} \in \Omega_{f_{\mathcal{N}_i}}$ should satisfy
\begin{subequations}\label{eq:terminal set conditions}
    \begin{align}
        x_{f_{\mathcal{N}_i}} &\in \mathcal{X}_{\mathcal{N}_i} = \bigtimes_{j \in \mathcal{N}_i} \mathcal{X}_{j},\\
        K_{f_i} x_{f_{\mathcal{N}_i}} & \in \mathcal{U}_i,
    \end{align}
\end{subequations}
where $ \mathcal{X}_{\mathcal{N}_i}  := \left\lbrace x_{\mathcal{N}_i} \in \mathbb{R}^{n} : \bar{a}_{il}^{\top} x_{\mathcal{N}_i} \leq \bar{d}_{il}, l=1,\ldots,n_{r_{\mathcal{N}_i}} \right\rbrace $.

\begin{theorem}\label{theorem:Omega_fi bounds}
	For each agent $ i \in \mathcal{M} $, if there exist scalars $ \phi_{ijl} \geq 0 $, $ \forall j \in \mathcal{N}_i $, $ l = 1,\ldots,n_{r_{\mathcal{N}_i}} $, and $ \psi_{ijp} \geq 0 $, $ \forall j \in \mathcal{N}_i $, $ p = 1,\ldots,m_{r_i} $ such that
	\begin{subequations}\label{eq:Omega_f1 bounds-X}
		\begin{align}
			& \begin{bmatrix}
				\displaystyle \sum_{j \in \mathcal{N}_i} \phi_{ijl} F_{ij} && (\alpha_{\mathcal{N}_i}^{\frac{1}{2}})^{\top} \bar{a}_{il} \\
				\star && \bar{d}_{il}
			\end{bmatrix} \succeq 0,\label{eq:Omega_f1 bounds-X-1}\\
			& \sum_{j \in \mathcal{N}_i} \phi_{ijl} \leq \bar{d}_{il},\;l=1,\ldots,n_{r_{\mathcal{N}_i}},\label{eq:Omega_f1 bounds-X-2}
		\end{align}
	\end{subequations}
	and
	\begin{subequations}\label{eq:Omega_f1 bounds-U}
		\begin{align}
			& \begin{bmatrix}
			\displaystyle \sum_{j \in \mathcal{N}_i} \psi_{ijp} F_{ij} && (\alpha_{\mathcal{N}_i}^{\frac{1}{2}})^{\top} K_{f_i}^{\top} h_{ip} \\
			\star && g_{ip}
			\end{bmatrix} \succeq 0, \label{eq:Omega_f1 bounds-U-1}\\
			& \sum_{j \in \mathcal{N}_i} \psi_{ijp} \leq g_{ip},\;p=1,\ldots,m_{r_i},\label{eq:Omega_f1 bounds-U-2}
		\end{align}
	\end{subequations}
	then the conditions in \eqref{eq:terminal set conditions} are satisfied.
\end{theorem}

\begin{pf}
	See Appendix~\ref{appendix:proof-Omega_fi bounds}.
\end{pf}

Moreover, by using the Schur complement, the constraint~\eqref{eq:MPC terminal constraint} can be rewritten as 
\begin{equation}\label{eq:Omega_fi terminal constraint}
	\begin{bmatrix}
		\alpha_{i}^{\frac{1}{2}}  && \bar{x}_i(N)^{\top}\\
		\star && \alpha_{i}^{\frac{1}{2}} F_{i}^{-1}
	\end{bmatrix} \succeq 0, \forall  i \in \mathcal{M}.
\end{equation}

To this end, the conditions for the constraint~\eqref{eq:MPC terminal constraint} in~\eqref{problem:RDMPC} can be summarised as follows:
\begin{equation*}
	 \bar{x}_i(N) \in \Omega_{f_i} (\alpha_i) \Leftrightarrow \eqref{eq:Omega_fi condition}, \eqref{eq:Omega_f1 bounds-X}, \eqref{eq:Omega_f1 bounds-U},\eqref{eq:Omega_fi terminal constraint},
\end{equation*}
for given $ F_i $, $ K_{f_i} $. The decision variables are $ \alpha_{i}^{\frac{1}{2}} $, $ \sigma_i $, $ \sigma_{ij} $, $ \phi_{ijl} $ and $ \psi_{ijl} $. Besides, $ \alpha_{\mathcal{N}_i}^{\frac{1}{2}} = T_{\mathcal{N}_i} \alpha T_{\mathcal{N}_i}^{\top} $ and $ \alpha^{\frac{1}{2}} = \mathrm{diag}(\alpha_1^{\frac{1}{2}} I_{n_1},\ldots, \alpha_M^{\frac{1}{2}} I_{n_M}) $.

\subsection{Summary of robust DMPC algorithm}\label{section:summary algorithm}

The proposed robust DMPC algorithm has both offline (synthesis) and online (optimisation) components. These are summarised in Algorithm~\ref{algorithm:offline} and~\ref{algorithm:online}, respectively. 

\begin{algorithm}
	\caption{Offline synthesis of $ K/K_{\mathcal{N}_i} $, $ K_{f_i} $ and $ P_{f_i} $} 
	\label{algorithm:offline}
	\begin{algorithmic}[1]
		\State Solve the optimisation problem~\eqref{problem:K synthesis} to obtain $ K $ or alternatively use conditions in Corollary~\ref{corollary:distributed RPI set}-\ref{corollary:Kni nonempty sets} to obtain $ K_{\mathcal{N}_i} $.
		\For{$t=0,\ldots,N-1$}
			\State Compute the set $ \mathcal{R}(t) $ based on~\eqref{eq:set definition R-Q}.
			\State Compute the tightened constraint sets $ \bar{\mathcal{X}}(t) $ and $ \bar{\mathcal{U}}(t) $ based on~\eqref{eq:tightened constraints}.
			\State Project the global constraint sets into local ones $  \bar{\mathcal{X}}_{\mathcal{N}_i}(t)  $ and $ \bar{\mathcal{U}}_i(t) $ based on~\eqref{eq:tightened local constraints}.
		\EndFor
		\State Obtain $ K_{f_i} $ and $ P_{f_i} $ for each agent $ i \in \mathcal{M} $.
	\end{algorithmic} 
\end{algorithm}

\begin{algorithm}
	\caption{Online robust DMPC}
	\label{algorithm:online} 
	\begin{algorithmic}[1]
		\State Choose and fix $ F_i = P_{f_i} $ and $ K_{f_i} $ from the offline synthesis for each agent $ i \in \mathcal{M} $.
		\While {$k \geq 0$}
			\State Each agent $ i \in \mathcal{M} $ measures its local current state $ x_i(k) $.
			\State Solve optimisation problem~\eqref{problem:RDMPC} with conditions~\eqref{eq:Omega_fi condition}, \eqref{eq:Omega_f1 bounds-X}, \eqref{eq:Omega_f1 bounds-U}, \eqref{eq:Omega_fi terminal constraint} for terminal constraints by distributed optimisation, where agents $ \mathcal{N}_i $ iteratively communicate.
			\State Apply $ u_i(k) = \kappa_{N} (x_i(k)) $ as in~\eqref{eq:RDMPC control action}.
		\EndWhile
	\end{algorithmic} 
\end{algorithm}

\section{Simulation results}\label{section:simulation}

\begin{figure}[thbp]
	\centering
	\includegraphics[width=\hsize]{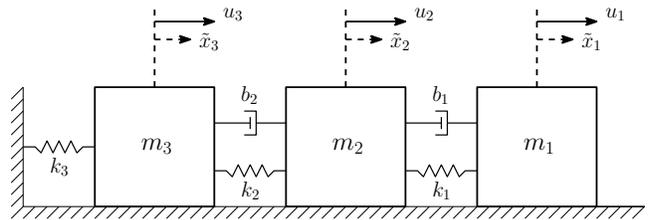}
	\caption{The chain of three masses, connected by springs and dampers.}
	\label{fig:mass springer damper}
\end{figure}

In this section, we use a mass-spring-damper system to demonstrate the proposed robust DMPC. Let us consider the system as shown in Fig.~\ref{fig:mass springer damper}.

Let the vectors of states and inputs be selected as $ x =  \left[ \tilde{x}_1, \dot{\tilde{x}}_1, \tilde{x}_2, \dot{\tilde{x}}_2, \tilde{x}_3, \dot{\tilde{x}}_3 \right]^{\top}\in \mathbb{R}^{6}$ and $ u = \left[u_1,u_2,u_3 \right]^{\top} \in \mathbb{R}^3 $. The continuous-time state-space model with additive disturbances can be formulated as follows:
\begin{align*}
	\dot{x} &= \begin{bsmallmatrix}
		0 & 1 & 0 & 0 & 0 & 0\\
		-\frac{k_1}{m_1} & -\frac{b_1}{m_1} & \frac{k_1}{m_1} & \frac{b_1}{m_1} & 0 & 0\\
		0 & 0 & 0 & 1 & 0 & 0\\
		\frac{k_1}{m_2} & \frac{b_1}{m_2} & -\frac{k_1+k_2}{m_2} & -\frac{b_1 + b_2}{m_2} & \frac{k_2}{m_2} & \frac{b_2}{m_2}\\
		0 & 0 & 0 & 0 & 0 & 1\\
		0 & 0 & \frac{k_2}{m_3} & \frac{b_2}{m_3} & -\frac{k_2+k_3}{m_3} & -\frac{b_2}{m_3}
	\end{bsmallmatrix} x \\
	& \quad + \begin{bsmallmatrix}
		0 & 0 & 0\\
		\frac{1}{m_1} & 0 & 0\\
		0 & 0 & 0\\
		0 &	\frac{1}{m_2} & 0\\
		0 & 0 & 0\\
		0 & 0 & \frac{1}{m_3}
	\end{bsmallmatrix} u + \begin{bmatrix}
		w_1\\w_2\\w_3
	\end{bmatrix}.
\end{align*}
where the parameters of masses, spring constants and damping coefficient are chosen asymmetrically with $ k_1 > k_2 > k_3 $ and $ b_1 > b_2 $ from $ m_1,m_2,m_3 \in \left[ 5,10\right] \mathrm{kg} $, $ k_1,k_2,k_3 \in \left[ 0.8,1.2 \right] \mathrm{N}\cdot \mathrm{m} $ and $ b_1,b_2 \in \left[ 0.8,2 \right] \mathrm{kg}/\mathrm{s} $, respectively. The resulting discrete-time LTI system in the form of~\eqref{eq: global linear system} can be obtained by using the Euler discretisation method to the above continuous-time state-space model with the sampling time $ T_s= 0.1 \mathrm{s} $. Each mass $ i \in \mathcal{M} := \left\lbrace  1,2,3 \right\rbrace  $ with the external force $ u_i $ can be considered as one agent. Asymmetric constraints are imposed as 
\begin{align*}
	\mathcal{X}_1 &= \left\lbrace x_1 \in \mathbb{R}^2: \left[-10,-10 \right]^{\top} \leq x_1 \leq \left[10,10 \right]^{\top}\right\rbrace,\\
	\mathcal{X}_2 &= \left\lbrace x_2 \in \mathbb{R}^2: \left[-2,-3 \right]^{\top} \leq x_2 \leq \left[2,3 \right]^{\top}\right\rbrace,\\
	\mathcal{X}_3 &= \left\lbrace x_3 \in \mathbb{R}^2: \left[-3,-5 \right]^{\top} \leq x_3 \leq \left[3,5 \right]^{\top}\right\rbrace,\\
	\mathcal{U} &= \left\lbrace u \in \mathbb{R}^3: \left|u_1 \right| \leq 10,\left|u_2 \right| \leq 1.5, \left|u_3 \right| \leq 5 \right\rbrace,
\end{align*}
and the disturbances are unknown but bounded in given sets 
\begin{align*}
	\mathcal{W}_1 &= \left\lbrace w_1 \in \mathbb{R}^2: \left[ -0.15,-0.3\right] \leq w_1 \leq \left[ 0.15,0.3\right]^{\top} \right\rbrace,\\
	\mathcal{W}_2 &= \left\lbrace w_2 \in \mathbb{R}^2: \left[ -0.05,-0.1\right] \leq w_2 \leq \left[ 0.05,0.1\right]^{\top} \right\rbrace,\\
	\mathcal{W}_3 &= \left\lbrace w_3 \in \mathbb{R}^2: \left[ -0.05,-0.1\right] \leq w_3 \leq \left[ 0.05,0.1\right]^{\top} \right\rbrace.
\end{align*}

\begin{table}[thbp]
	\caption{Parameters for robust DMPC.} \label{table:parameters}
	\vspace{0.2cm}
	\begin{center}
		\begin{tabular}{c c c}
			\toprule
			&$Q_{\mathcal{N}_i} $  & $ R_{i} $   \\
			\midrule
		   Agent 1  & $\mathrm{diag}(10,10)$  & 0.1 \\
		   Agent 2 & $ \mathrm{diag}(1,1) $  & 0.01  \\
		   Agent 3 & $ \mathrm{diag}(2.5,2.5) $  & 0.05  \\
			\bottomrule
		\end{tabular}
	\end{center}
\end{table}

\begin{table*}[thbp]
	\caption{Synthesis results for robust DMPC.} \label{table:offline synthesis}
	\vspace{0.2cm}
	\begin{center}
		\begin{tabular}{c c c c}
			\toprule
			& $K_{\mathcal{N}_i} $ & $ K_{f_i} $ & $ P_{f_i}  $ \\
			\midrule
			 Agent 1 & $ \begin{bmatrix}
				-0.29 & -0.88 & -0.67 & -0.82
			\end{bmatrix} $  &  $ \begin{bmatrix}-7.68 & -11.99 & -0.26 & -0.33]\end{bmatrix} $ &  $\begin{bmatrix}
			151.92 & 57.76\\
			57.76 & 92.08
			\end{bmatrix} $ \\
			Agent 2 & $ \begin{bmatrix} -0.57 & -0.70 & -0.29 & -0.83 & -0.70 & -0.28\end{bmatrix} $  & $ \begin{bmatrix}-0.29 & -0.38 & -12.39 & -19.07 & -0.11 & -0.17\end{bmatrix} $ & $\begin{bmatrix}
			25.95 &  14.23\\
			14.23 &  21.53
			\end{bmatrix}$ \\
			Agent 3 &  $ \begin{bmatrix} -0.41& -1.21 & -1.01 &  -3.17\end{bmatrix} $  & $ \begin{bmatrix} -0.12 7 -0.19 & -9.69 & -15.30\end{bmatrix} $  & $\begin{bmatrix}
				68.39 &  40.33\\
				40.33 &  62.28
			\end{bmatrix} $\\
			\bottomrule
		\end{tabular}
	\end{center}
\end{table*}

The weighting matrices for the stage cost functions are given in Table~\ref{table:parameters}. 

\begin{figure}[t]
	\centering
	\subfigure[Case 1]{\includegraphics[width=\hsize]{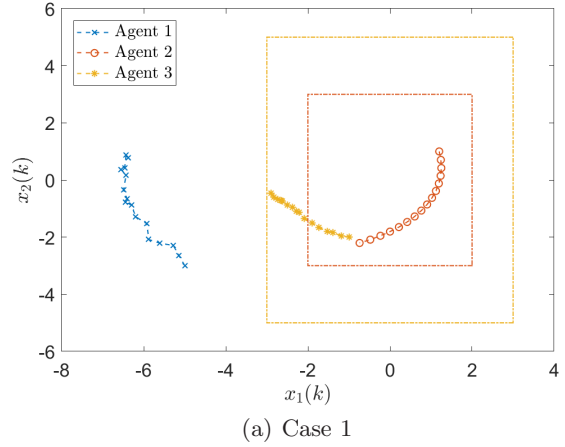}}
	\subfigure[Case 2]{\includegraphics[width=\hsize]{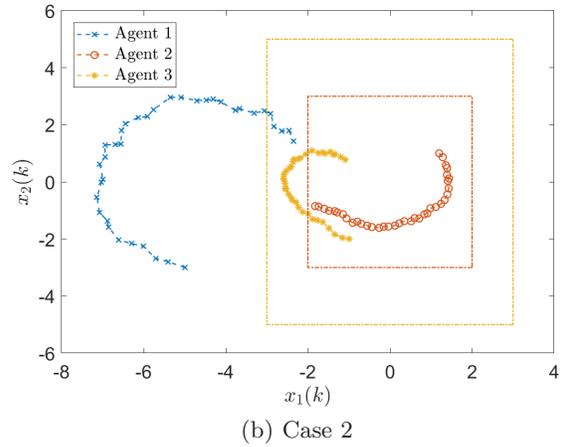}}
	\caption{State trajectories under nominal DMPC with noise sequences leading to infeasibility.}
	\label{fig:infeasible results}
\end{figure}

\begin{figure}[thbp]
	\centering
	\includegraphics[width=\hsize]{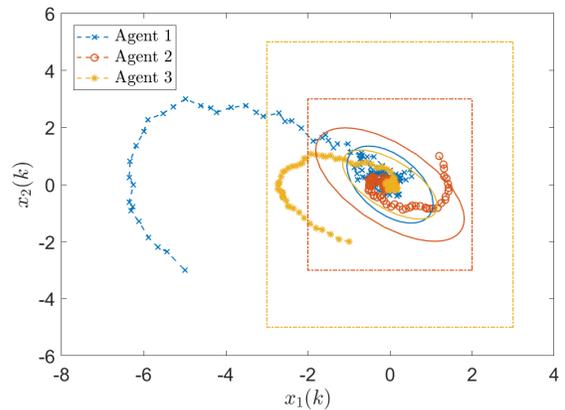}
	\caption{State trajectories under robust DMPC, and local terminal sets.}
	\label{fig:closed-loop state trajectories}
\end{figure}

\begin{figure}[thbp]
	\centering
	\includegraphics[width=\hsize]{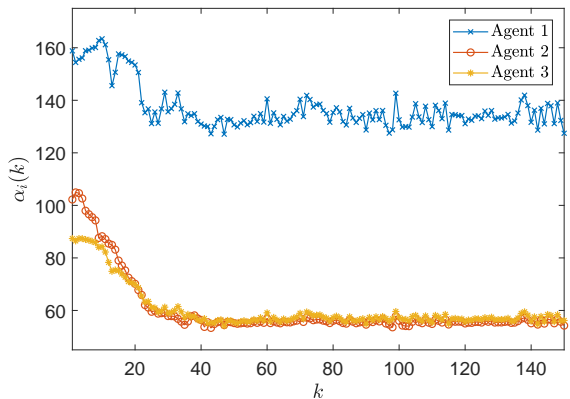}
	\caption{The optimal $ \alpha_i(k) $.}
	\label{fig:alpha}
\end{figure}

For comparison, in the closed-loop simulations, we have implemented the nominal DMPC proposed in~\cite{Darivianakis2020} with this uncertain system. The initial condition $ x(0) = [-5,-3,1.2,1,-1,-2]^{\top} $ is given and the prediction horizon is considered as $ N=5 $. With a number of realisations of the disturbance sequences sampled from $ \mathcal{W}_1 \times \mathcal{W}_2 \times \mathcal{W}_3 $, two sequences leading to infeasible closed-loop results were identified and denoted 'Case 1' and 'Case 2'. The closed-loop state trajectories of three agents are shown in Fig.~\ref{fig:infeasible results}. 

The same disturbance sequences are now applied to the closed-loop system with the proposed robust DMPC controller. Offline implementation of Algorithm~\ref{algorithm:offline}, leads to the local feedback gains and terminal cost matrices along with the terminal feedback gains shown in Table~\ref{table:offline synthesis}. The online implementation of Algorithm~\ref{algorithm:online} then allows the closed-loop trajectories to be determined for the same disturbance sequences that led to infeasibility with the nominal controller. As illustrated in Fig.~\ref{fig:closed-loop state trajectories}, the feasibility of the closed-loop system is now retained, thereby validating the results of Theorem~\ref{theorem:recursive feasibility}. Similarly, the convergence of the system towards the origin demonstrates that the desired ISS property is achieved, thereby validating Theorem~\ref{theorem:ISS stability}. This stability is further illustrated by plotting the trajectories of the $ \alpha_i  $ parameters in Fig.~\ref{fig:alpha}, leading to the ellipsoidal terminal sets shown in Fig.~\ref{fig:closed-loop state trajectories}.


\section{Conclusions}\label{section:conlcusion}

In this paper, we have proposed a robust DMPC formulation for discrete-time LTI systems subject to disturbances. The closed-loop system with the proposed robust DMPC controller has been proved to be recursively feasible and ISS stable in the presence of unknown-but-bounded disturbances. We have presented algorithms for the offline (synthesis) problem as well as the online (optimisation) problem that have provable guarantees. As a future direction, the proposed robust DMPC formulation can be extended into tracking DMPC as well as economic DMPC.

\begin{ack}                               
	We would like to thank Prof. Ren\'{e} Boel from the Ghent University for providing insight on distributed model predictive control. We also would like to thank Prof. Iman Shames from The University of Melbourne for useful discussions on distributed optimisation techniques.
\end{ack}

\appendix
\section{Proofs in Section~\ref{section:analysis}}

\subsection{Proof of Theorem~\ref{theorem:recursive feasibility}}\label{appendix:recursive feasibility}

The feasible solutions of~\eqref{problem:RDMPC} at any time step $ k \geq 0 $ are denoted as in~\eqref{eq:RDMPC feasible solutions}. The control action at time step $ k $ is chosen to be $  u_i(k) = \kappa_{N} (x_i(k)) $ as in~\eqref{eq:RDMPC control action}. Due to the constraint~\eqref{eq:MPC initialisation}, the control action $  u_i(k) = \kappa_{N} (x_i(k)) $ also satisfies~\eqref{eq:tube-based control action}.

According to the constraints~\eqref{eq:MPC constraint-xbar}-\eqref{eq:MPC constraint-ubar}, the corresponding closed-loop state $ x_i(k) $ and $ u_i(k) $ are feasible at time step $ k $, that is, $ x_i(k) \in \mathcal{X}_i $ and $ u_i(k) \in \mathcal{U}_i $ when $ t=0 $. After applying the optimal control action~\eqref{eq:RDMPC control action}, we can obtain $ x_{i}(k+1) $ from the system~\eqref{eq: global linear system}.

Referring to~\eqref{eq:tube-based control action}, with $ x_i(k+1) $, $ \forall i \in \mathcal{M} $ and feasible solution at time step $ k $, we define a sequence of shifted nominal control inputs
\begin{equation}\label{eq: shifted inputs-1}
	\begin{split}
		& \bar{u}_i(t; x_i(k+1)) := \bar{u}_{i}^{*}(t+1; x_{i}(k)) \\
		& \qquad + L_i K (\bar{x} (t; x(k+1)) - \bar{x}^{*}(t+1; x(k)) ),
	\end{split}
\end{equation}
for $ t=0,\ldots,N-2 $, $ \forall i \in \mathcal{M} $, and when $ t=N-1 $,
\begin{equation}\label{eq: shifted inputs-2}
\bar{u}_i(N-1; x_i(k+1)) := K_{f_i} \bar{x}_{\mathcal{N}_i} (N-1; x_i(k+1)).
\end{equation}

We can also define the shifted error for the global system states as
\begin{align}\label{eq:shift error}
\bar{e}(t) := \bar{x}(t; x(k+1)) - \bar{x}^{*}(t+1; x(k)),
\end{align}
with 
\begin{align*}
\bar{e}(0) &= \bar{x}(0; x(k+1)) - \bar{x}^{*}(1; x(k))\\
&= x(k+1) - \mathrm{col}_{i \in \mathcal{M}} \left\lbrace  A_{\mathcal{N}_{i}} x_{\mathcal{N}_{i}}(k) + B_{i} \kappa_{N} (x_i(k)) \right\rbrace  \\
&= w(k) \in \mathcal{W}. 
\end{align*}

With~\eqref{eq: shifted inputs-1}, the shifted error dynamics can be described as 
\begin{equation}\label{eq:shift error dynamics}
\begin{split}
\bar{e}(t+1) &=  \mathrm{col}_{i \in \mathcal{M}} \left\lbrace \left(   A_{\mathcal{N}_{i}} T_{\mathcal{N}_i} + B_{i} U_i K  \right)  \bar{e}(t) \right\rbrace \\
&= A_{K} \bar{e}(t) .
\end{split}
\end{equation}

From $ \bar{e}(0) $ and~\eqref{eq:shift error dynamics}, we denote shifted error sets $ \bar{e}(t) \in \bar{\mathcal{E}}(0) = \mathcal{W} $, and $ \bar{e}(t) \in \bar{\mathcal{E}}(t) = A_{K}^{t} \mathcal{W} $, $ t \geq 1 $, $ \forall i \in \mathcal{M} $.

We now check the feasibility of all the constraints in~\eqref{problem:RDMPC} at time step $ k+1 $, $ \forall i \in \mathcal{M} $.
\begin{itemize}
	\item State constraint~\eqref{eq:MPC constraint-xbar}: for $ t=0,1,\ldots,N-1 $, 
	\begin{align*}
	\bar{x}_{\mathcal{N}_i}(t; x_i(k+1)) & \in  \bar{x}_{\mathcal{N}_i}^{*}(t+1; x_{i}(k)) \oplus T_{\mathcal{N}_i} \bar{\mathcal{E}}(t) \\
	& \subseteq \bar{\mathcal{X}}_{\mathcal{N}_i}(t+1)  \oplus T_{\mathcal{N}_i} \bar{\mathcal{E}}(t) \\
	&= T_{\mathcal{N}_i} \mathcal{X} \ominus T_{\mathcal{N}_i}  \left\lbrace \mathcal{R}(t+1) \ominus \bar{\mathcal{E}}(t) \right\rbrace \\
	&= T_{\mathcal{N}} \mathcal{X}_{i} \ominus T_{\mathcal{N}_i} \mathcal{R}(t)\\
	&= \bar{\mathcal{X}}_{\mathcal{N}_i}(t). 
	\end{align*}
	\item Input constraint~\eqref{eq:MPC constraint-ubar}: for $ t=0,1,\ldots,N-1 $,
	\begin{align*}
	\bar{u}_i(t; x_i(k+1)) &\in \bar{u}_{i}^{*}(t+1; x_{i}(k)) \oplus L_i K \bar{\mathcal{E}}(t) \\
	&\subseteq \bar{\mathcal{U}}_i(t+1) \oplus L_i K \bar{\mathcal{E}}(t) \\
	&= L_i \mathcal{U}_i \ominus  L_i K \left\lbrace \mathcal{R}(t+1) \ominus \bar{\mathcal{E}}(t) \right\rbrace \\
	&= L_i \mathcal{U}_i \ominus  L_i K  \mathcal{R}(t)\\
	&= \bar{\mathcal{U}}_i(t).
	\end{align*} 
	\item Terminal constraint~\eqref{eq:MPC terminal constraint}: Since
	\begin{align*}
	\bar{x}_i(N-1; x_i(k+1)) &\in \bar{x}_{i}^{*}(N; x_{i}(k)) \oplus T_i \bar{\mathcal{E}}(N-1)   \\
	&\subseteq \Omega_{f_i} (\alpha_i) \oplus T_{i} \bar{\mathcal{E}}(N-1),
	\end{align*}
	and then by using the terminal control law~\eqref{eq: shifted inputs-2}, when $ t=N $, it holds $  \bar{x}_i(N; x_i(k+1)) \in  \Omega_{f_i} (\alpha_i)  $ by the condition~\eqref{eq:robust adaptive termianl set} in Definition~\ref{definition:terminal sets}.
	\item Initial condition constraint~\eqref{eq:MPC initialisation}: $ \bar{x}_i(0; x_i(k+1))  =  x_i(k+1) $.
\end{itemize}

Thus, the optimisation problem~\eqref{problem:RDMPC} is also feasible at time step $ k+1 $, $ \forall k \geq 0 $. 	\qed

\subsection{Proof of Theorem~\ref{theorem:ISS stability}}\label{appendix:ISS stability}

Denote the optimal cost of~\eqref{problem:RDMPC} at any time step $ k \geq 0 $ as $ V_N^*(x(k)) := \sum_{i \in \mathcal{M}} V_{N_i}^*(x_i(k)) $. Since $ V_N^*(x(k)) $ is positive-definite and continuous in a neighbourhood of the coordinate origin, there exist $ \mathcal{K} $ functions $ \beta_{1} $ and $ \beta_{2} $ such that $ \beta_1 (\left\| x(k) \right\| ) \leq V_N^*(x(k)) \leq \beta_2 (\left\| x(k) \right\| ) $. Then, at time step $ k+1 $, let us define
\begin{align*}
\Delta V_N := &V_N(x(k+1))- V_N^*(x(k)) \\
= &\sum_{i \in \mathcal{M}} \left( V_{N_i}(x_i(k+1)) - V_{N_i}^*(x_i(k)) \right),
\end{align*}
where for $ i \in \mathcal{M} $,
\begin{align*}
&V_{N_i}(x_i(k+1)) - V_{N_i}^*(x_i(k)) \\
=& V_{f_i}(\bar{x}_i(N;x_i(k+1))) - V_{f_i}^*(\bar{x}_i(N;x_i(k)))  \\
& +\sum_{t=0}^{N-1}  \left( \ell_i (\bar{x}_{\mathcal{N}_i}(t;x_i(k+1)),\bar{u}_i(t;x_i(k+1)) \right)  \\
&  - \sum_{t=0}^{N-1}  \left( \ell_i (\bar{x}_{\mathcal{N}_i}^*(t;x_i(k)),\bar{u}_i^*(t;x_i(k)) \right)\\
=& \|\bar{x}_i(N;x_i(k+1))\|_{P_{f_i}}^2 - \| \bar{x}_i(N;x_i(k))\|_{P_{f_i}}^2  \\
&  +\sum_{t=0}^{N-1}  \left( \left\| \bar{x}_{\mathcal{N}_i}(t;x_i(k+1)) \right\|_{Q_{\mathcal{N}_i}}^{2} + \left\| \bar{u}_i(t;x_i(k+1)) \right\|_{R_{i}}^{2} \right)  \\
&  - \sum_{t=0}^{N-1}  \left( \left\| \bar{x}_{\mathcal{N}_i}^*(t;x_i(k)) \right\|_{Q_{\mathcal{N}_i}}^{2} + \left\| \bar{u}_i^*(t;x_i(k)) \right\|_{R_{i}}^{2} \right) .
\end{align*}

For $ t = 0,1,\ldots, N-2 $, from~\eqref{eq:shift error}-\eqref{eq:shift error dynamics}, we have
\begin{align*}
\bar{x}_{\mathcal{N}_i}(t; x_i(k+1)) - \bar{x}_{\mathcal{N}_i}^{*}(t+1; x_{i}(k))
&= T_{\mathcal{N}_i}\bar{e}(t)\\
&= T_{\mathcal{N}_i} A_{K}^{t} w(k).
\end{align*}

Then, it follows
\begin{align*}
& \left\| \bar{x}_{\mathcal{N}_i}(t; x_i(k+1)) \right\|_{Q_{\mathcal{N}_i}}^2 -\left\|  \bar{x}_{\mathcal{N}_i}^{*}(t+1; x_{i}(k))  \right\|_{Q_{\mathcal{N}_i}}^2 \\
= &  \| A_{K}^{t} w(k) \|_{\bar{Q}_{i}}^2 + 2 (T_{\mathcal{N}_i} ^{\top} Q_{\mathcal{N}_i} \bar{x}_{\mathcal{N}_i}^{*}(t+1; x_{i}(k)) )^{\top} A_{K}^{t} w(k)  \\
\leq & \| A_{K}^{t} w(k) \|_{\bar{Q}_{i}}^2 + 2 \| T_{\mathcal{N}_i} ^{\top} Q_{\mathcal{N}_i} \bar{x}_{\mathcal{N}_i}^{*}(t+1; x_{i}(k)) \| \| A_{K}^{t} w(k) \|\\
\leq & \| A_{K}^{t} w(k) \|_{\bar{Q}_{i}}^2 + 2 c_{i,1} \| A_{K}^{t} w(k) \|,
\end{align*}
where $ \bar{Q}_{i} = T_{\mathcal{N}_i}^{\top} Q_{\mathcal{N}_i} T_{\mathcal{N}_i} $, and $ c_{i,1} $ is an upper bound of $ \| T_{\mathcal{N}_i}^{\top}  Q_{\mathcal{N}_i} \bar{x}_{\mathcal{N}_i} \| $ for given matrices $ T_{\mathcal{N}_i} $ and $ Q_{\mathcal{N}_i}  $, $ \forall \bar{x}_{\mathcal{N}_i} \in T_{\mathcal{N}_i} \mathcal{X} $.

Similarly, due to~\eqref{eq: shifted inputs-1}, we  have
\begin{align*}
&\bar{u}_{i}(t; x_i(k+1)) - \bar{u}_{i}^{*}(t+1; x_{i}(k)) \\= &L_i K \bar{e}(t)
= L_i K A_{K}^{t} w(k).
\end{align*}

Then, it follows
\begin{align*}
& \left\| \bar{u}_{i}(t; x_i(k+1)) \right\|_{R_{i}}^2 -\left\|  \bar{u}_{i}^{*}(t+1; x_{i}(k))  \right\|_{R_{i}}^2 \\
\leq & \| A_{K}^{t} w(k)  \|_{\bar{R}_i}^2 + 2 c_{i,2} \| A_{K}^{t} w(k) \|,
\end{align*}
where $ \bar{R}_{i} = K^{\top} L_i^{\top} R_i L_i K $, and $ c_{i,2} $ is an upper bound of $ \| K^{\top} L_i^{\top} R_i \bar{u}_{i} \| $ for given matrices $ K $, $ L_i $ and $ R_{i}  $, $ \forall u_i \in L_i \mathcal{U} $.

When $ t=N-1 $, we can obtain $ \bar{x}_{\mathcal{N}_i}(N-1; x_i(k+1)) - \bar{x}_{\mathcal{N}_i}^{*}(N; x_{i}(k)) = T_{\mathcal{N}_i}  A_{K}^{N-1} w(k) $. Besides, based on~\eqref{eq: shifted inputs-2}, we can derive 
\begin{align*}
&\left\| \bar{x}_{\mathcal{N}_i}(N-1; x_i(k+1)) \right\|_{Q_{\mathcal{N}_i}}^2 + \left\| \bar{u}_{i}(N-1; x_i(k+1)) \right\|_{R_i}^2\\
=& \left\| \bar{x}_{\mathcal{N}_i}(N-1; x_i(k+1)) \right\|_{Q_{\mathcal{N}_i}+K_{f_i}^{\top} R_i K_{f_i}}^2\\
\leq & \left\| \bar{x}_{\mathcal{N}_i}^{*}(N; x_i(k)) \right\|_{Q_{\mathcal{N}_i}+K_{f_i}^{\top} R_i K_{f_i}}^2 \\
& + \| A_{K}^{N-1} w(k) \|_{\bar{Q}_{f_i}}^2 + 2 c_{i,3} \|A_{K}^{N-1} w(k)\|,
\end{align*}
where $ \bar{Q}_{f_i} = T_{\mathcal{N}_i}^{\top} \left(  Q_{\mathcal{N}_i}+K_{f_i}^{\top} R_i K_{f_i}\right) T_{\mathcal{N}_i}  $,  and $ c_{i,3} $ is an upper bound of $ \| T_{\mathcal{N}_i}^{\top}(Q_{\mathcal{N}_i}+ K_{f_i}^{\top} R_i K_{f_i}) \bar{x}_{\mathcal{N}_i} \|, $ for given matrices $ T_{\mathcal{N}_i} $, $ Q_{\mathcal{N}_i}  $, $ R_i $ and $ K_{f_i} $, $ \forall \bar{x}_{\mathcal{N}_i} \in T_{\mathcal{N}_i} \mathcal{X} $.

Based on a shifted control input from~\eqref{eq: shifted inputs-2} at the prediction step $ N $, we know $ \bar{x}_i^{*}(N+1;x_i(k)) = A_{K_{f_i}}  \bar{x}_{\mathcal{N}_i}^{*}(N;x_i(k)) $, which gives 
\begin{align*}
\| \bar{x}_i^{*}(N+1;x_i(k)) \|_{P_{f_i}} ^2= \| \bar{x}_{\mathcal{N}_i}^{*}(N;x_i(k))  \|_{A_{K_{f_i}}^{\top} P_{f_i}  A_{K_{f_i}}} ^2.
\end{align*}

Also based on shifted control inputs from~\eqref{eq: shifted inputs-1} and~\eqref{eq: shifted inputs-2} at the prediction step $ N $, it also comes 
\begin{align*}
\bar{x}_{i}(N;x_i(k+1)) - \bar{x}_{i}^{*}(N+1;x_i(k)) = A_{K_{f_i}} T_{\mathcal{N}_i}  A_{K}^{N-1} w(k).
\end{align*}

Therefore, we have
\begin{align*}
& \| \bar{x}_i(N;x_i(k+1)) \|_{P_{f_i}} ^2 \\
\leq & \| \bar{x}_{\mathcal{N}_i}^{*}(N;x_i(k))  \|_{A_{K_{f_i}}^{\top} P_{f_i}  A_{K_{f_i}}} ^2 + \|  A_{K}^{N-1} w(k) \|_{\bar{P}_{f_i}}^2 \\
& \; + 2 c_{i,4} \|  A_{K}^{N-1} w(k) \|,
\end{align*}
where $ \bar{P}_{f_i} = T_{\mathcal{N}_i}^{\top} A_{K_{f_i}}^{\top} P_{f_i} A_{K_{f_i}} T_{\mathcal{N}_i} $, and $ c_{i,4} $ is an upper bound of $ \| T_{\mathcal{N}_i}^{\top}   A_{K_{f_i}}^{\top} P_{f_i} x_i \| $ for given matrices $ K_{f_i} $ and $ P_{f_i}  $.

From the condition~\eqref{eq:negativity of local terminal cost}, we have
\begin{align*}
& \| \bar{x}_{\mathcal{N}_i}^{*}(N;x_i(k))  \|_{A_{K_{f_i}}^{\top} P_{f_i}  A_{K_{f_i}}} ^2  - \| \bar{x}_i^{*}(N;x_i(k)) \|_{P_{f_i}} ^2\\
\leq & -\left\| \bar{x}_{\mathcal{N}_i}^{*}(N; x_i(k)) \right\|_{Q_{\mathcal{N}_i}+K_{f_i}^{\top} R_i K_{f_i}}^2 + \gamma_i (\bar{x}_{\mathcal{N}_i}^{*}(N;x_i(k))) .
\end{align*}

As a result, we thus obtain
\begin{align*}
& V_{N_i}(x_i(k+1)) - V_{N_i}^*(x_i(k))  \\
\leq &  \gamma_i (\bar{x}_{\mathcal{N}_i}^{*}(N;x_i(k))) +\lambda_i ( \| w_i(k) \|) \\
&  - \| \bar{x}_{\mathcal{N}_i}^{*}(0;x_i(k)) \|_{Q_{\mathcal{N}_i}}^{2} - \| \bar{u}_i^*(0;x_i(k)) \|_{R_{i}}^{2}
\end{align*}
where 
\begin{align*}
\lambda_i ( \| w(k) \|) = & \sum_{t=0}^{N-2} \Big(  \| A_{K}^{t} w(k) \|_{\bar{Q}_{i}}^2 + 2 c_{i,1} \| A_{K}^{t} w(k) \| \\
& + \| A_{K}^{t} w(k)  \|_{\bar{R}_i}^2 + 2 c_{i,2} \| A_{K}^{t} w(k) \| \Big)\\
& + \| A_{K}^{N-1} w(k) \|_{\bar{Q}_{f_i}}^2 + 2 c_{i,3} \|A_{K}^{N-1} w(k)\|\\
& + \|  A_{K}^{N-1} w(k) \|_{\bar{P}_{f_i}}^2  + 2 c_{i,4} \|  A_{K}^{N-1} w(k) \|,
\end{align*}

By optimality, we know $ V_{N_i}^{*}(x_i(k+1) \leq V_{N_i}(x_i(k+1) $. Then, by proceed with sum, we can obtain with~\eqref{eq:total gamma}
\begin{align*}
& V_N^*(x(k+1))- V_N^*(x(k)) \\
\leq & \sum_{i \in \mathcal{M}} \left( - \| \bar{x}_{\mathcal{N}_i}^{*}(0;x_i(k)) \|_{Q_{\mathcal{N}_i}}^{2}  + \lambda_i ( \| w(k) \|) \right),
\end{align*}
which is an ISS-Lyapunov function as stated in~[Definition 7 ]\cite{limon2009input}.  Thus, the closed-loop system is ISS stable. \qed

\section{Proofs in Section~\ref{section:synthesis}}
\subsection{Proof of Theorem~\ref{theorem:RPI set}}\label{appendix:RPI ellipsoid}

The condition of the RPI set can be written as $ x(k+1) \in \mathcal{Z} $, $ \forall x(k) \in \mathcal{Z}$ and $\forall w \in \mathcal{W}$. By using the S-procedure~\cite[Chapter 2.6.3]{Boyd-LMI1994}, the above condition is satisfied if there exist $ \tau_1 \geq 0 $, $ \tau_2 \geq 0 $ such that
\begin{align*}
	&\begin{bmatrix}
		- A_{K}^{\top} P A_{K}& -A_{K}^{\top} P & 0\\
		-P A_{K} & -P & 0\\
		0 & 0 & 1
	\end{bmatrix} 
	-\tau_1 \begin{bmatrix}
		- P & 0 & 0\\
		0 & 0 & 0\\
		0 & 0 & 1
	\end{bmatrix}\\
	& \qquad -\tau_2 \begin{bmatrix}
		0 & 0 & 0\\
		0 & -W & 0\\
		0 & 0 & 1
	\end{bmatrix}\succeq 0
\end{align*}

The above condition is equivalent to~\eqref{eq:synthesis K-2} and 
\begin{align*}
	\begin{bmatrix}
	\tau_1 P &0\\
	0 & \tau_2 W 
	\end{bmatrix}
	- \begin{bmatrix}
		A_{K}^{\top}\\I
	\end{bmatrix} P \begin{bmatrix}
		A_{K} & I
	\end{bmatrix}\succeq 0.
\end{align*}

By using the Schur complement to the above condition, we can obtain
\begin{align*}
	\begin{bmatrix}
		 \tau_1 P & 0 & A_{K}^{\top}\\
		 0 & \tau_2 W & I \\
		 A_{K} & I & P^{-1}
	\end{bmatrix} \succeq 0.
\end{align*}

Again, by using the Schur complement, we have 
\begin{align*}
	\begin{bmatrix}
		 \tau_2 W & I \\
		I & P^{-1}
	\end{bmatrix} 
	-\begin{bmatrix}
		0 \\A_K
	\end{bmatrix} (\tau_1 P)^{-1} \begin{bmatrix}
		0 & A_{K}^{\top}
	\end{bmatrix}\succeq 0.
\end{align*}

By using the Schur complement to the above condition with setting $ Y  =  K P^{-1} $ and $ S =P^{-1} $, we thus obtain~\eqref{eq:synthesis K-1}. \qed

\subsection{Proof of Theorem~\ref{theorem:K nonempty sets}} \label{appendix:K nonempty sets}

First, let us discuss the condition to guarantee $ \mathcal{X} \ominus \mathcal{Z} $. By using the results in~\cite[Chapter 8.4.2]{boyd2004convex}, the condition to guarantee $ \mathcal{Z} \subseteq \mathcal{X} $ can be expressed as
\begin{align*}
	\left\| P^{-\frac{1}{2}} a_j\right\| \leq d_j ,\; j = 1,\ldots,n_r.
\end{align*}
which can be rewritten as
\begin{align*}
	d_j^2 - a_j^{\top} P^{-1} a_j \geq 0, \; j = 1,\ldots,n_r.
\end{align*}

By using the Schur complement to the above condition with setting $ Y  =  K P^{-1} $ and $ S =P^{-1} $, we can obtain~\eqref{eq:synthesis K-3}.

Then, similarly, considering $ \mathcal{U} $ defined in~\eqref{eq:convex sets U} and the control gain $ K $, we have $  h_{j}^{\top} K x \leq g_j $, $ j=1,\ldots,m_{r}  $. The condition to guarantee non-empty $ \mathcal{U} \ominus K \mathcal{Z} $ can be expressed as
\begin{align*}
	g_j^2 - h_j^{\top} P^{-1} h_j \geq 0, \; j = 1,\ldots,m_r.
\end{align*}

By using the Schur complement to the above condition, we thus obtain~\eqref{eq:synthesis K-4}. \qed

\subsection{Proof of Corollary~\ref{corollary:distributed RPI set}}\label{appendix:proof-corollary 1}

Similar to the proof of Theorem~\ref{theorem:RPI set}, for each agent $ i $, the condition $ x_i(k+1) \in \mathcal{Z}_i$, $ \forall x_i(k) \in \mathcal{Z}_i $ and $ \forall w_i \in \mathcal{W}_i $ is satisfied if there exist two scalars $ \bar{\tau}_i \geq 0 $, $ \bar{\tau}_{ij} \geq 0 $, $ \forall j \in \mathcal{N}_i $ such that~\eqref{eq:synthesis Kni-2} and
\begin{align*}
	\begin{bmatrix}
	\displaystyle \sum_{j \in \mathcal{N}_i} \bar{\tau}_{ij} P_{ij} &0\\
	0 & \bar{\tau}_i W_i
	\end{bmatrix}
	- \begin{bmatrix}
	A_{K_{\mathcal{N}_i}}^{\top}\\I
	\end{bmatrix} P_i \begin{bmatrix}
	A_{K_{\mathcal{N}_i}} & I
	\end{bmatrix}\succeq 0,
\end{align*}
with $ A_{K_{\mathcal{N}_i}} := A_{\mathcal{N}_i} + B_i K_{\mathcal{N}_i} $. By using the Schur complement twice, the above condition is equivalent to
\begin{align*}
	\begin{bmatrix}
	\bar{\tau}_i W_i & I \\
	I & P_i^{-1}
	\end{bmatrix} 
	-\begin{bmatrix}
	0 \\A_{K_{\mathcal{N}_i}}
	\end{bmatrix} \left( \displaystyle \sum_{j \in \mathcal{N}_i} \bar{\tau}_{ij} P_{ij} \right) ^{-1} \begin{bmatrix}
	0 & A_{K_{\mathcal{N}_i}}^{\top}
	\end{bmatrix}\succeq 0.
\end{align*}

Pre-multiplying and post-multiplying~\eqref{eq:synthesis Kni-1} by $ \left[ I \;\; -\varTheta_i\right]  $ and $ \left[ I \;\; -\varTheta_i \right]^{\top}  $ with $ \varTheta_i^{T} = \left[ 0 \;\; A_{K_{\mathcal{N}_i}}^{\top}\right] $ can obtain the above condition with setting $ S_i = P_i^{-1} $ and $ Y_i = K_{\mathcal{N}_i} G_i $. \qed

\subsection{Proof of Theorem~\ref{theorem:Omega_fi}} \label{appendix:proof-Omega_fi}

From the condition~\eqref{eq:robust adaptive termianl set}, we have 
\begin{align*}
	A_{K_{f_i}} (x_{\mathcal{N}_i} + e_{\mathcal{N}_i}) \in \Omega_{f_i} (\alpha_{i}), \forall x_j \in \Omega_{f_j} (\alpha_{j}), \forall j \in \mathcal{N}_i,
\end{align*}
and $ \forall e_ {\mathcal{N}_i} \in \bar{\mathcal{E}}_{\mathcal{N}_i} (N-1) $, which is equivalent to
\begin{align*}
	& (x_{\mathcal{N}_i} + e_{\mathcal{N}_i})^{\top} A_{K_{f_i}}^{\top} F_{i} A_{K_{f_i}} (x_{\mathcal{N}_i} + e_{\mathcal{N}_i})  \leq \alpha_{i},\\
	& \quad \mathrm{for \; all} \;x_j^{\top} F_i x_j \leq \alpha_j , \forall j \in \mathcal{N}_i, \; \mathrm{and}\; e_{\mathcal{N}_i}^{\top} E_{\mathcal{N}_i} e_{\mathcal{N}_i} \leq 1.
\end{align*}

Refer to~\cite[Proposition 2]{Darivianakis2020}, we set $ x_i = \alpha_{i}^{\frac{1}{2}} s_i $, $ x_{\mathcal{N}_i} = \alpha_{\mathcal{N}_i}^{\frac{1}{2}} s_{\mathcal{N}_i} $ and $ e_{\mathcal{N}_i} = \alpha_{\mathcal{N}_i}^{\frac{1}{2}} z_{\mathcal{N}_i} $.

The above condition is equivalent to
\begin{align*}
	&(s_{\mathcal{N}_i} + z_{\mathcal{N}_i})^{\top} (A_{K_{f_i}}\alpha_{\mathcal{N}_i}^{\frac{1}{2}} )^{\top} F_{i} A_{K_{f_i}} \alpha_{\mathcal{N}_i}^{\frac{1}{2}}  (s_{\mathcal{N}_i} + z_{\mathcal{N}_i})  \leq \alpha_{i},\\
	& \quad \mathrm{for \; all} \;s_{\mathcal{N}_i}^{\top} F_{ij} s_{\mathcal{N}_i}\leq 1, \forall j \in \mathcal{N}_i, \\
	& \quad \mathrm{and}\; z_{\mathcal{N}_i}^{\top} (\alpha_{\mathcal{N}_i}^{\frac{1}{2}} )^{\top} E_{\mathcal{N}_i} \alpha_{\mathcal{N}_i}^{\frac{1}{2}} z_{\mathcal{N}_i} \leq 1.
\end{align*}

By using the S-procedure~\cite[Chapter 2.6.3]{Boyd-LMI1994}, the above condition is satisfied if there exist scalars $ \sigma_i \geq 0 $, $ \sigma_{ij} \geq 0 $, $ \forall j \in \mathcal{N}_i $, $ \forall i \in \mathcal{M} $ such that~\eqref{eq:Omega_fi condition-2} and
\begin{align*}
	& \begin{bmatrix}
		\displaystyle\sum_{j \in \mathcal{N}_i} \sigma_{ij} F_{ij} & 0\\
		0 & (\alpha_{\mathcal{N}_i}^{\frac{1}{2}} )^{\top} \sigma_i E_{\mathcal{N}_i} \alpha_{\mathcal{N}_i}^{\frac{1}{2}}
	\end{bmatrix}\\
	& \quad -\begin{bmatrix}
		(A_{K_{f_i}}\alpha_{\mathcal{N}_i}^{\frac{1}{2}} )^{\top}\\(A_{K_{f_i}}\alpha_{\mathcal{N}_i}^{\frac{1}{2}} )^{\top}
	\end{bmatrix}
	\alpha_{i}^{-\frac{1}{2}} F_{i} \begin{bmatrix}
		A_{K_{f_i}}\alpha_{\mathcal{N}_i}^{\frac{1}{2}} && A_{K_{f_i}}\alpha_{\mathcal{N}_i}^{\frac{1}{2}}
	\end{bmatrix} \succeq 0.
\end{align*}

By using the Schur complement and arranging its rows and columns, we can obtain
\begin{align*}
	\begin{bmatrix}
		\displaystyle\sum_{j \in \mathcal{N}_i} \sigma_{ij} F_{ij} && ( \alpha_{\mathcal{N}_i}^{\frac{1}{2}} )^{\top} A_{K_{f_i}}^{\top} &&  0 \\
		\star && \alpha_{i}^{\frac{1}{2}} F_{i}^{-1} && A_{K_{f_i}}\alpha_{\mathcal{N}_i}^{\frac{1}{2}}\\
		\star && \star && (\alpha_{\mathcal{N}_i}^{\frac{1}{2}})^{\top} \sigma_i E_{\mathcal{N}_i} \alpha_{\mathcal{N}_i}^{\frac{1}{2}}
	\end{bmatrix} \succeq 0.
\end{align*}

Again, by using the Schur complement twice, we thus obtain~\eqref{eq:Omega_fi condition-1}. \qed

\subsection{Proof of Theorem~\ref{theorem:Omega_fi bounds}} \label{appendix:proof-Omega_fi bounds}

Also based on the results in~\cite[Chapter 8.4.2]{boyd2004convex}, the condition for $ \Omega_{f_i} (\alpha_i)$ satisfying $ \mathcal{X}_{\mathcal{N}_i} = \bigtimes_{j \in \mathcal{N}_i} \mathcal{X}_{j}$ can be formulated as 
\begin{align*}
	& \left\| \bar{a}_{il}^{\top}x_{\mathcal{N}_i} \right\|  \leq \bar{d}_{il},\;\mathrm{for \; all} \;x_j^{\top} F_i x_j \leq \alpha_j , \forall j \in \mathcal{N}_i,
\end{align*}
for each $  l=1,\ldots,n_{r_{\mathcal{N}_i}} $. 

Set $ x_i = \alpha_{i}^{\frac{1}{2}} s_i $, $ x_{\mathcal{N}_i} = \alpha_{\mathcal{N}_i}^{\frac{1}{2}} s_{\mathcal{N}_i}$. The above condition is equivalent to
\begin{align*}
	& s_{\mathcal{N}_i}^{\top} (\alpha_{\mathcal{N}_i}^{\frac{1}{2}} )^{\top} \bar{a}_{il} \bar{d}_{il}^{-1} \bar{a}_{il}^{\top}\alpha_{\mathcal{N}_i}^{\frac{1}{2}}  s_{\mathcal{N}_i}  \leq \bar{d}_{il},\\
	& \quad \mathrm{for \; all} \;s_{\mathcal{N}_i}^{\top} F_{ij} s_{\mathcal{N}_i} \leq 1, \forall j \in \mathcal{N}_i,
\end{align*}
which is satisfied if there exist scalars $ \phi_{ijl} \geq 0 $, $ \forall j \in \mathcal{N}_i $ such that~\eqref{eq:Omega_f1 bounds-X-2} and 
\begin{align*}
	\displaystyle \sum_{j \in \mathcal{N}_i} \phi_{ijl} F_{ij}  - (\alpha_{\mathcal{N}_i}^{\frac{1}{2}} )^{\top} \bar{a}_{il}\bar{d}_{il}^{-1} \bar{a}_{il}^{\top}\alpha_{\mathcal{N}_i}^{\frac{1}{2}}  \succeq 0.
\end{align*}

By using the Schur complement to the above condition, we thus obtain~\eqref{eq:Omega_f1 bounds-X-1}.

On the other hand, following the same procedure, we have
\begin{align*}
	& x_{\mathcal{N}_i}^{\top} K_{f_i} ^{\top} h_{ip} g_{ip}^{-1} h_{ip}^{\top} K_{f_i} x_{\mathcal{N}_i}  \leq g_{ip},\\
	& \quad \mathrm{for \; all} \;x_j^{\top} F_i x_j \leq \alpha_j , \forall j \in \mathcal{N}_i,
\end{align*}
for each $  p=1,\ldots,m_{r_i} $, is satisfied, if the conditions in~\eqref{eq:Omega_f1 bounds-U} hold. \qed

\bibliographystyle{apacite}        
\bibliography{autosam}           



\end{document}